\documentclass[aps,amsfonts,amssymb,amsmath]{revtex4}
\usepackage{epsf}

\begin{document}
\title{Degeneracy breaking and intervalley scattering due to short-ranged
impurities in finite single-wall carbon nanotubes}
\author{Edward McCann and Vladimir I. Fal'ko}
\affiliation{Department of Physics, Lancaster University,
Lancaster, LA1 4YB, United Kingdom}

\begin{abstract}
We present a theoretical study of degeneracy breaking due to short-ranged
impurities in finite, single-wall, metallic carbon nanotubes.
The effective mass model is used to describe the slowly varying spatial
envelope wavefunctions of spinless electrons near the Fermi level at two
inequivalent valleys (K-points) in terms of the four component Dirac equation
for massless fermions, with the role of spin assumed by pseudospin due to
the relative amplitude of the wave function on the sublattice atoms
(``A'' and ``B'').
Using boundary conditions at the ends of the tube that neither
break valley degeneracy nor mix pseudospin eigenvectors,
we use degenerate perturbation theory to show that the presence of impurities
has two effects. Firstly, the position of the impurity with respect to the
spatial variation of the envelope standing waves results in a sinusoidal
oscillation of energy level shift as a function of energy.
Secondly, the position of the impurity within the hexagonal
graphite unit cell produces a particular $4\times 4$ matrix structure of the
corresponding effective Hamiltonian.
The symmetry of this Hamiltonian with respect to pseudospin flip is related
to degeneracy breaking and, for an armchair tube, the symmetry with respect
to mirror reflection in the nanotube axis is related to pseudospin mixing.
\end{abstract}

\maketitle
%
\section{Introduction}

Much of the interest in carbon nanotubes has been
motivated by the desire to develop new nanoscale electrical
devices \cite{saito98,dek99}.
The electronic properties of nanotubes follow from the band structure of
a two-dimensional sheet of graphite which is a semi-metal,
having a vanishing energy gap at the six corners, K-points, of the hexagonal
first Brillouin zone. A single-wall nanotube may be thought of as a graphene
sheet rolled up to form a nanometre-diameter cylinder. Periodicity around
the circumference results in quantized transverse wavevectors leading to
metallic or semiconducting behaviour depending on whether the K-point
wavevector ${\bf K}$ is an allowed wavevector.

A finite nanotube should possess discrete energy levels corresponding to standing
waves typical of a confined quantum particle. Evidence of discrete levels
was seen in transport measurements \cite{bock97,tans97} a few years ago,
followed by the direct observation of sinusoidal standing wave patterns by
scanning tunneling microscopy \cite{ven99,lem01}
with wavevectors corresponding to those near the K-point ${\bf K}$.
More recently, Coulomb blockade measurements on carbon nanotube quantum
dots \cite{lia02,bui02,cob02} have seen varying degrees of evidence for the
fourfold periodicity of shell filling that would be in agreement with
expectations based on the spin and valley (K-point) structure.

In this paper, we will consider the interplay between two sources of valley
degeneracy breaking in a finite nanotube, namely isolated impurities
and the boundaries themselves.
The aim is to show how the character of an impurity determines
the extent of valley degeneracy breaking, resulting in the possibility to
observe either twofold or fourfold periodicity of shell filling \cite{cob02}.
As far as boundaries are concerned, a number of authors
\cite{rubio99,roche99,wu00,y+a01,jiang02,m+f04}
have modelled finite-length nanotubes in order to describe the atomic scale
variation of standing waves patterns and the opening of an energy gap
displaying an oscillating dependence on the tube length.
For impurities, theoretical studies of open nanotubes by Ando and
co-workers \cite{ando98a,ando98b,ando99} have shown that short-ranged
potentials (typical range smaller than the lattice constant of graphite)
produce back-scattering, but not long-ranged potentials.
For an armchair tube, it was demonstrated that impurities preserving mirror
reflection in the nanotube axis do not mix the bonding $\pi$ and antibonding
$\pi^{\ast}$ energy bands \cite{ando99,choi00,song02}.
For closed nanotubes, a recent density-functional calculation \cite{ke03}
has shown how a small number of defects may reduce the four-fold
periodicity of shell filling to two-fold.

In the scanning tunneling microscopy measurements of Ref.~\cite{lem01} an
additional slow spatial modulation of the standing waves was observed. It
was interpreted as being a beating envelope function with wavevector ${\bf q}$,
$\left| {\bf q}\right| \ll \left| {\bf K}\right| $, resulting from
the interference of left and right moving waves with slightly different total
wavevectors ${\bf K}\pm {\bf q}$. The effective mass
model \cite{d+m84,a+a93,k+m97,mceuen99} provides an analytical
description of the electronic structure near the K point where the total
wavevector is ${\bf k}={\bf K}+{\bf q}$ and the dispersion relation is linear
$E= sv\left| {\bf q}\right| $, $v$ is the Fermi velocity and $s = \pm 1$
for the conduction and valence band, respectively. For spinless
electrons, the envelope wavefunction $\Psi \left( {\bf q},{\bf r}\right)$ has four
components corresponding to two inequivalent atomic sites in the hexagonal
graphite lattice (``A'' and ``B'') and to two inequivalent K-points in the
hexagonal first Brillouin zone.
The resulting eigenvalue equation for $\Psi$ is the massless Dirac equation
written in the ``chiral'' or ``spinor'' representation,
\begin{eqnarray}
-iv{\bf \alpha .\nabla }\Psi  &=&E\Psi ;\hspace{0.5in}\alpha =\left( 
\begin{array}{cc}
{\bf \sigma } & 0 \\ 
0 & -{\bf \sigma }
\end{array}
\right) ; \label{diracintro} \\
{\bf \sigma } &=&e^{i\eta \sigma _{z}/2}\left( \sigma _{x}{\bf \hat{\imath}}
+\sigma _{y}{\bf \hat{\jmath}}\right) e^{-i\eta \sigma _{z}/2} , \nonumber
\end{eqnarray}
where the role of spin is assumed by the relative
amplitudes on the A and B atomic sites (``pseudospin''): ${\bf \sigma }$ is a
vector in the
$(x,y)$ plane rotated by the chiral angle $\eta$ of the tube.
Also, $v=\left( \sqrt{3}/2\right) a\gamma$ is the Fermi velocity,
$a$ is the lattice constant of graphite and $\gamma$
is the nearest neighbour transfer integral.

Since we are interested in perturbations of a clean nanotube that may
destroy valley degeneracy, we must identify the symmetry that preserves
degeneracy.
The pseudospin of a 2D graphite sheet does not transform in the same way as
real spin because certain transformations result in
a swapping of the orientation of A and B atoms.
This leads us to identify an operator $\rho_{z}$ that
flips pseudospin but commutes with the clean effective
Hamiltonian, Eq.~(\ref{diracintro}),
\begin{eqnarray}
\rho_{z} =\left( 
\begin{array}{cc}
0 & i \sigma_z\\ 
i \sigma_z & 0
\end{array}
\right) . \label{sz}
\end{eqnarray}
In general, the two degenerate eigenvectors $\{ \Psi_1 , \Psi_2 \}$
corresponding to the two non-equivalent K-points of the Dirac equation for
a clean, metallic nanotube may be labelled using the component of
pseudospin along the tube axis $\Sigma_{a}$
or using pseudo-helicity $\lambda$.
Therefore, the pseudospin-flip operator $\rho_{z}$ relates the degenerate
eigenvectors to each other,
$\rho_{z} \Psi_1 \rightarrow \Psi_2$.
We may make two statements about the consequence of the symmetry of
a particular perturbation $\delta H$.
The first is that perturbations that are symmetric
in the pseudospin-flip operator $\rho_{z}^{-1} \delta H \rho_{z} = \delta H$
preserve pseudospin and do not break valley degeneracy.
Secondly, a perturbation that breaks pseudospin-flip symmetry,
$\rho_{z}^{-1} \delta H \rho_{z} \neq \delta H$, but is still symmetric
with respect to the operator $\Sigma_{a}$ measuring pseudospin
$\Sigma_{a}^{-1} \delta H \Sigma_{a} = \delta H$,
will break degeneracy without mixing the pseudospin eigenvectors.
Since pseudospin is the relative amplitude of the wavefunction on the
A and B atomic sites, a given perturbation must differentiate between
adjacent atoms in order to break pseudospin symmetry.
In other words, the influence of the perturbation must vary spatially
on the scale of the graphite lattice constant $a$: such a perturbation
is described as being short-ranged.

\begin{figure}[t]
\centerline{\epsfxsize=0.3\hsize
\epsffile{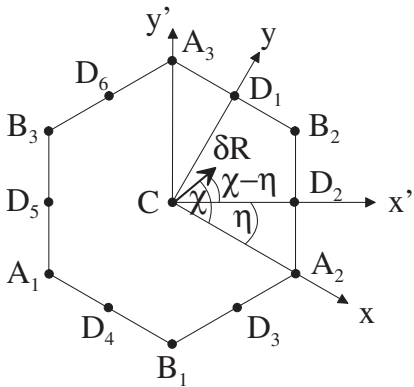}}
\caption{\label{fig:positions}
The positions with respect to the graphite unit cell of the perturbing
potential are labelled as A, B, C, and D.
Carbon atomic positions are at the six corners of the hexagon,
there are three A atomic positions $\{ A_1, A_2, A_3 \}$
and three B atomic positions $\{ B_1, B_2, B_3 \}$.
We also consider the potential to be near the centre of the unit cell (C) or
at one of six positions half-way between neighbouring atoms ($D_1$ to $D_6$).
An additional small deviation ${\bf \delta R}$ of the potential position
is shown (greatly exaggerated) for the C position, with direction
described by angle $\chi$ in the nanotube coordinates $(x,y)$.
The figure has chiral angle $\eta = \pi /6$ corresponding to an
armchair tube.
}
\end{figure}

We will investigate how a perturbing short-range potential breaks the
inter-valley degeneracy.
The position of a potential within the hexagonal graphite unit cell
will produce a specific $4 \times 4$ matrix structure of the resulting
effective Hamiltonian, and the symmetry of the matrix will determine
the extent of degeneracy breaking.
As the ultimate limit of a short-range potential,
we consider a delta function potential because it simplifies the
calculations and the resulting analysis.
We would like to stress that our intent is not to produce exact
quantitative results that describe the influence of impurities,
but to characterise possible symmetry breaking properties.
The positions of the potential we consider are shown with relation to the
hexagonal graphite unit cell in Fig.~\ref{fig:positions}.
They are near an A type atomic site, labelled A in the figure,
near a B type atomic site, labelled B,
near the centre of the unit cell, labelled C,
or near the half-way point between neighbouring atoms, labelled D.

%
%
%
%

The paper is organised as follows.
In Section~\ref{S:emm}, the effective mass model leading to the Dirac
equation is briefly described along with a discussion of its symmetry
properties. Section~\ref{S:bc} is an introduction to the boundary
conditions of a closed carbon nanotube and the
resulting energy spectrum of a clean nanotube is calculated.
In Section~\ref{S:dpt}, we use degenerate perturbation theory to show how
valley degeneracy is broken by a short-range potential and to examine the
relationship between the position of the potential and symmetry.
In Appendix~B we give a brief account of a non-perturbative
calculation of the energy spectrum for the example of an impurity exactly
on an atomic site.

\section{Symmetry properties of the effective mass model\label{S:emm}}

In the effective mass model of two-dimensional graphite \cite{d+m84},
the total wavefunction $\Psi_{\rm tot}$\ is written as a
linear combination of four components $m = \left\{ 1,2,3,4\right\}$
corresponding to two K-points $\mu = \left\{ 1,2\right\}$ and $\pi$-type
atomic orbitals $\varphi_{j}({\bf r}-{\bf R_j})$ on two non-equivalent atomic
sites $j = \left\{ A,B\right\}$ in the unit cell,
\begin{equation}
\Psi_{\rm tot}\left( {\bf r}\right) =
\sum_{m=1}^{4}\left\{ \Phi _{m}^{(0)}\left( {\bf r}\right) - {\bf G}_{m}
\left( {\bf r}\right) .{\bf \nabla }+\ldots \right\}
\psi_{m}\left( {\bf r}\right) ,  \label{totalwf}
\end{equation}
where
\begin{eqnarray}
\Phi _{m}^{(0)}\left( {\bf r}\right) &=&
\frac{1}{\sqrt{N}}\sum_{{\bf R}_{{\bf j}}}^{N} e^{i{\bf K_{\mu }.R}_{{\bf j}}}
\, \varphi_{j} ({\bf r}-{\bf R}_{{\bf j}}),  \label{phi0} \\
{\bf G}_{m}\left( {\bf r}\right) &=&
\frac{1}{\sqrt{N}}\sum_{{\bf R}_{{\bf j}}}^{N}e^{i{\bf K_{\mu }.R}_{{\bf j}}}
\, \varphi_{j}({\bf r}-{\bf R}_{{\bf j}})
({\bf r}-{\bf R}_{{\bf j}}),  \label{gm}
\end{eqnarray}
are Bloch type functions constructed from the atomic orbitals,
${\bf R}_{{\bf j}}$ is the position of an atom in real space and the
summation is over the number of unit cells $N\gg 1$.
The functions $\psi_{m}\left( {\bf r}\right)$ are components of the
envelope function $\Psi\left( {\bf q},{\bf r}\right)$. Substituting
this expression for $\Psi_{tot}$\ into the Schr\"{o}dinger equation
and integrating with respect to fast degrees of freedom that vary on the
scale of the unit cell leads to the Dirac equation Eq.(\ref{diracintro})
for the envelope function $\Psi$.
We label the two non-equivalent K-points as $K$ and $\widetilde{K}$
with wavevectors ${\bf K} = \left( \pm 4 \pi / 3a , 0 \right)$,
and the components of $\Psi$
are written in the order $KA$, $KB$, $\widetilde{K}B$, $\widetilde{K}A$.
The appearance of the chiral angle of the tube $\eta$ in the Dirac
equation shows that the axes of the $(x^{\prime},y^{\prime})$ ``graphite''
coordinate system have been rotated to be transverse and parallel to the
tube axis, labelled $(x,y)$ in Fig.~\ref{fig:positions}.
Applying periodic boundary conditions to the wavefunction $\Psi_{tot}$,
Eq.(\ref{totalwf}), in the direction transverse to the nanotube axis
produces a condition for the envelope function $\Psi$ that leads to
metallic or semiconducting behaviour depending on whether the transverse
component of wavevector ${\bf q}$ is allowed to be zero \cite{a+a93,k+m97}.

In order to highlight the separate K-point space and AB space structure
present in carbon nanotubes we adopt a matrix direct product notation
using $\left\{ \sigma_{x},\sigma_{y},\sigma_{z},I_{\sigma }\right\}$
for $2\times 2$ Pauli matrices and the unit matrix that operate within a
block (`AB space') and
$\left\{ \Pi_{x},\Pi_{y},\Pi_{z},I_{\Pi }\right\}$ for $2\times 2$
Pauli matrices and the unit matrix that operate in K-point space.
For example, the operator $\alpha$ may be written as
$\alpha =\Pi_{z}\otimes {\bf \sigma }$, and
the usual operators for the reflection of real spin
in a plane that reverses the Cartesian coordinate $n = x$, $y$ or $z$ are
$\rho_n = i\Pi_{x}\otimes
e^{i\eta \sigma _{z}/2} \sigma_n e^{-i\eta \sigma _{z}/2}$.
As mentioned in the Introduction, the pseudospin of a 2D graphite sheet
does not transform in the same way as real spin because certain
transformations result in a swapping of the orientation of A and B atoms.
This additional operation is described by the ``pseudospin-flip''
operator, Eq.~(\ref{sz}), $\rho_{z} = \Pi_{x} \otimes i\sigma_z$ that corresponds
to a reflection of real spin in the $(x,y)$ plane.
For example, an active rotation of the 2D graphite sheet
anticlockwise by $\pi /3$ about the perpendicular $z$ axis,
$\Psi \left( {\bf r^{\prime}} \right) = C_6 \Psi \left( {\bf r} \right)$,
is described by
$C_6 = \rho_{z} R(\pi /3) = \Pi_{x} \otimes \exp \left( (2\pi i/3) \sigma_z \right)$
where $R(\theta ) = I_{\Pi} \otimes \exp \left( (i\theta /2) \sigma_z \right)$
is a continuous rotation operator.

We consider the nanotube axis to be parallel to the unit vector
${\bf {\hat n}}= (\sin \eta ,\cos \eta ,0)$
in the $(x^{\prime},y^{\prime})$ graphite coordinates,
so that it points along the $y$-axis in the $(x,y)$ nanotube coordinates,
Fig.~(\ref{fig:positions}).
In this rotated coordinate system, the component of the pseudospin operator
along the positive $y$-axis is $\Sigma_{a} = I_{\Pi} \otimes 
e^{i\eta \sigma _{z}/2} \sigma_{y} e^{-i\eta \sigma _{z}/2}$
and the pseudo-helicity operator is
$\lambda = \left| {\bf q} \right|^{-1} I_{\Pi} \otimes e^{i\eta \sigma _{z}/2}
\left( -i\sigma_{y}\partial_y \right) e^{-i\eta \sigma _{z}/2}$.
For an armchair tube, a mirror reflection across the nanotube axis
(the $y$-axis in Fig.~(\ref{fig:positions})) is accompanied by an exchange of A and
B atomic positions so that it is described not by operator $\rho_{x}$ but by the
combination $\rho_{z} \rho_{x}$ representing reflection of real spin accompainied
by an additional spin-flip.
It turns out that $\Sigma_{a} = i \rho_{z} \rho_{x}$, so we may draw
the conclusion that, for an armchair tube, the operator measuring
pseudospin also represents a mirror reflection across the nanotube axis.
The situation is different for a zigzag tube because reflection across the
nanotube axis (the $y^{\prime}$-axis in Fig.~(\ref{fig:positions}))
is not accompanied by an exchange of A and B atomic positions so that it is described
by operator $\rho_{x}$, not $\Sigma_{a} = i \rho_{z} \rho_{x}$.
This means that potential positions that are symmetric with respect to
mirror reflection across the axis of an armchair tube, such
as positions $D_1$ and $D_4$ in Fig.~(\ref{fig:positions}),
will also be symmetric with respect to the pseudospin operator
and will break degeneracy without mixing the pseudospin eigenvectors.
Since pseudospin is related to the underlying molecular orbital state
\cite{mceuen99}, this statement is equivalent to saying that impurities
preserving mirror reflection in the nanotube axis do not result in 
mixing of the bonding $\pi$ and antibonding
$\pi^{\ast}$ energy bands \cite{del98,mat98,ando99,choi00,song02}.
On the other hand, potential positions that are symmetric with respect to
mirror reflection across the axis of a zigzag tube, such
as positions $A_3$ and $B_1$ in Fig.~(\ref{fig:positions}),
will not be symmetric with respect to the pseudospin operator.

\section{Single particle energy spectrum of a closed nanotube}\label{S:bc}

In this section we calculate the form of non-interacting single particle
standing waves and the corresponding energy spectrum in a closed nanotube.
For simplicity, we will consider only metallic nanotubes
with arbitrary chiral angle $\eta $.
We suppose that the $x$ axis is perpendicular to the tube axis and we
consider only the zero momentum transverse mode so that
$|E| < 2 \pi v /L_c$ where $L_c = |{\bf C_h}|$ is the circumference.
The Dirac equation is diagonal in K-point space, so that, for
an open nanotube, there are two right moving ($\Psi _{K}^{(R)}$ and $\Psi
_{\widetilde{K}}^{(R)}$) and two left moving ($\Psi _{K}^{(L)}$ and $\Psi _{
\widetilde{K}}^{(L)}$) plane wave solutions:
\begin{eqnarray*}
\Psi _{K}^{(R)} &=&Ae^{iqy}\left(
\begin{array}{c}
1 \\ 
ise^{-i\eta } \\ 
0 \\ 
0
\end{array}
\right) ; \qquad \Psi _{K}^{(L)}=Be^{-iqy}\left( 
\begin{array}{c}
1 \\ 
-ise^{-i\eta } \\ 
0 \\ 
0
\end{array}
\right) ; \\
\Psi _{\widetilde{K}}^{(R)} &=&Ce^{iqy}\left( 
\begin{array}{c}
0 \\ 
0 \\ 
1 \\ 
-ise^{-i\eta }
\end{array}
\right) ; \qquad \Psi _{\widetilde{K}}^{(L)}=De^{-iqy}\left( 
\begin{array}{c}
0 \\ 
0 \\ 
1 \\ 
ise^{-i\eta }
\end{array}
\right) ,
\end{eqnarray*}
where $A$, $B$, $C$ and $D$ are arbitrary constants, $q$ is the wavevector
along the tube and we consider $q\geq 0$ and $E=svq$, $s=\pm 1$.
The solutions $\Psi_{K}^{(R)}$\ and $\Psi_{\widetilde{K}}^{(L)}$\ are
eigenvectors of pseudospin component $\Sigma_{a}$
with eigenvalue $+s$, whereas the solutions $\Psi _{\widetilde{K}}^{(R)}$ and
$\Psi _{K}^{(L)}$ have eigenvalue $-s$. 
Also, the solutions $\Psi_{K}^{(R)}$\ and $\Psi_{K}^{(L)}$ are
eigenvectors of pseudo-helicity $\lambda$ with eigenvalue $+s$,
whereas the solutions $\Psi _{\widetilde{K}}^{(R)}$ and
$\Psi _{\widetilde{K}}^{(L)}$ have eigenvalue $-s$.

Note that the Hamiltonian $H_{2d}$ given in Eq.~(\ref{diracintro}) is two dimensional,
but, by taking into account only the lowest transverse mode, it becomes one dimensional
$H_{1d}$ in a metallic tube:
\begin{eqnarray}
H_{2d} &=& v \Pi_z \otimes e^{i\eta \sigma _{z}/2}
\left( -i\sigma_{x}\partial_x -i\sigma_{y}\partial_y \right) e^{-i\eta \sigma _{z}/2} ,
\label{h2d} \\
H_{1d} &=& v \Pi_z \otimes e^{i\eta \sigma _{z}/2}
\left( -i\sigma_{y}\partial_y \right) e^{-i\eta \sigma _{z}/2} ,
\label{h1d}
\end{eqnarray}
The pseudospin part of the one dimensional Hamiltonian $H_{1d}$ may be
diagonalised using a unitary transformation,
${\widetilde{H}_{1d}} = {\cal U}^{-1} H_{1d} {\cal U}$ \cite{fn1}:
\begin{eqnarray}
{\cal U} &=& \frac{I_{\Pi}}{\sqrt{2}} \otimes e^{i\eta \sigma _{z}/2}
\left( \sigma_y + \sigma_z \right) e^{-i\eta \sigma _{z}/2} ,
\label{trans} \\
{\widetilde{H}_{1d}} &=& v \Pi_z \otimes 
\left( -i\sigma_{z}\partial_y \right) , \label{h1diag}
\end{eqnarray}
and the corresponding eigenvectors
${\widetilde{\Psi}}_{K/\widetilde{K}}^{(L/R)} =
{\cal U}^{-1} \Psi_{K/\widetilde{K}}^{(L/R)}$
are eigenvectors of $\sigma _{z}$ in pseudospin space so they have only one
non-zero component out of four.

Now we will briefly describe the effective boundary conditions
for the envelope function $\Psi$ in a closed carbon nanotube,
and refer the reader to Ref.~\cite{m+f04} for more details.
There it was shown that energy independent hard wall boundary
conditions for the Dirac equation may be expressed in general terms as 
\begin{equation}
\Psi =M\Psi ;\hspace{0.4in}M^{2}=1;\hspace{0.4in}\left\{ {\bf n}_{{\bf B}}
{\bf .\alpha },M\right\} =0,  \label{bcintro}
\end{equation}
where $M$ is an Hermitian, unitary $4\times 4$ matrix $M^{2}=1$ with the
constraint that it anticommutes with the operator
${\bf n}_{{\bf B}}{\bf .\alpha }$, proportional to the component of the current
operator normal to the interface, ${\bf n}_{{\bf B}}$ is the unit vector normal
to the interface.
There are four possible linear combinations of matrices
satisfying these constraints on $M$, which, assuming ${\bf n}_{{\bf B}}$ is
a vector confined to the $\left( x,y\right) $ plane, may be written in terms
of a small number of arbitrary parameters:
\begin{eqnarray}
M_{1} &=&\cos \Lambda \left( I_{\Pi }\otimes {\bf n}_{1}{\bf .\sigma }
\right) +\sin \Lambda \left( \Pi _{z}\otimes {\bf n}_{{\bf 2}}{\bf .\sigma }
\right) ,  \label{m1} \\
M_{2} &=&\cos \Upsilon \left( {\bf \nu }_{{\bf 1}}{\bf .\Pi }\otimes
I_{\sigma }\right) +\sin \Upsilon \left( {\bf \nu }_{{\bf 2}}{\bf .\Pi }
\otimes {\bf n}_{{\bf B}}{\bf .\sigma }\right) ,  \label{m2} \\
M_{3} &=&\cos \Omega \left( {\bf \nu }_{{\bf 2}}{\bf .\Pi }\otimes {\bf n}_{
{\bf B}}{\bf .\sigma }\right) +\sin \Omega \left( I_{\Pi }\otimes {\bf n}_{
{\bf 1}}{\bf .\sigma }\right) ,  \label{m3} \\
M_{4} &=&\cos \Theta \left( {\bf \nu }_{{\bf 1}}{\bf .\Pi }\otimes I_{\sigma
}\right) +\sin \Theta \left( \Pi _{z}\otimes {\bf n}_{{\bf 2}}{\bf .\sigma }
\right) ,  \label{m4}
\end{eqnarray}
where the angles $\Lambda $,$\Upsilon $,$\Theta $ and $\Omega $ are
arbitrary, ${\bf n}_{{\bf 1}}$ and ${\bf n}_{{\bf 2}}$ are three-dimensional
space-like vectors satisfying the constraints ${\bf n}_{{\bf 1}}{\bf .n}_{
{\bf B}}={\bf n}_{{\bf 2}}{\bf .n}_{{\bf B}}={\bf n}_{{\bf 1}}.{\bf n}_{{\bf 
2}}=0$, and ${\bf \nu }_{{\bf 1}}$ and ${\bf \nu }_{{\bf 2}}$ are
two-dimensional (confined to the $\left( x,y\right) $ plane) space-like
vectors satisfying the constraint ${\bf \nu }_{{\bf 1}}{\bf .\nu }_{{\bf 2}
}=0$.

In principle, there are different ways of combining the right and
left moving plane waves in order to create standing waves.
The first possibility is that waves at the same K-point combine, namely
$\Psi_{K}^{(R)}$ and $\Psi_{K}^{(L)}$ form a standing wave with pseudo-helicity
eigenvalue $+s$, and $\Psi_{\widetilde{K}}^{(R)}$ and
$\Psi_{\widetilde{K}}^{(L)}$ form a standing wave with pseudo-helicity
eigenvalue $-s$. This situation is realised by the matrix $M_1$,
Eq.(\ref{m1}), because it is diagonal in K-point space.
A second possibility is that waves from opposite K-points combine,
namely $\Psi_{K}^{(R)}$ and $\Psi_{\widetilde{K}}^{(L)}$ form a standing wave
with pseudospin component eigenvalue $+s$, and $\Psi _{\widetilde{K}}^{(R)}$
and $\Psi _{K}^{(L)}$ form a standing wave with pseudospin component
eigenvalue $-s$.
This situation is realised by the matrix $M_2$, Eq.(\ref{m2}), because it is
off-diagonal in K-point space.
A third possibility is a combination of the previous two, with waves
scattered back at the boundary into a mixture of both of the K-points.
This situation is realised by the matrices $M_3$, Eq.(\ref{m3}),
and $M_4$, Eq.(\ref{m4}), because they have both diagonal and
off-diagonal in K-point space parts.

In the graphite coordinate system, we define the normal to the boundary ${\bf n}_{{\bf B}}$
in terms of the chiral angle of the tube $\eta $ and we choose two mutually orthogonal 3D vectors
${\bf n_1}$ and ${\bf n_2}$, and two additional orthogonal 2D vectors ${\bf \nu_1}$ and ${\bf \nu_2}$:
\begin{eqnarray}
{\bf n}_{{\bf B}}&=& (\sin \eta ,\cos \eta ,0), \nonumber \\
{\bf n_1}&=& \left(\cos \eta \sin \zeta ,-\sin\eta\sin\zeta ,\cos\zeta\right), \nonumber \\
{\bf n_2}&=& \left( \cos \eta \cos \zeta , -\sin\eta\cos\zeta,-\sin\zeta\right), \nonumber \\
{\bf \nu_1}&=& \left( \cos \xi ,\sin \xi, 0\right), \nonumber \\
{\bf \nu_2}&=& \left( -\sin \xi ,\cos \xi ,0\right), \nonumber 
\end{eqnarray}
This introduces two new mixing angles, $\zeta$ and $\xi$:
the arbitrary parameters contained within the boundary conditions
describe the amount of mixing between different discrete symmetries.
Table~1 shows a summary of the discrete symmetries of the boundary conditions
in terms of the orientation of the vectors ${\bf n_1}$, ${\bf n_2}$,
${\bf \nu_1}$ and ${\bf \nu_2}$.
In addition to $\rho_{z}$ and $\Sigma_a$ we consider parity
$P =  \Pi_{x} \otimes I_{\sigma}$, corresponding to a rotation by $\pi$ about
the $z$ axis ($x \rightarrow -x$ and $y \rightarrow - y$),
and charge conjugation ($C$) that involves the
complex conjugation operator combined with $C = -i \Pi_y \otimes \sigma_y$.
The angles $\zeta$ and $\xi$ mix terms with different symmetry with respect
to $\rho_{z}$: $\zeta = 0$ and $\xi = 0$ correspond to evenness with respect to
$\rho_{z}$ whereas $\zeta = \pi /2$ and $\xi = \pi /2$ correspond to oddness.
Since pseudospin and/or pseudo-helicity label different states at the same
energy, values of $\zeta$ and $\xi$ not equal to multiples of $\pi /2$ will
lead to broken degeneracy.
The angles $\Lambda $,$\Upsilon $,$\Theta $ and $\Omega $ mix different
symmetries with respect to combinations of $P$, $C$ and $\rho_{z}$.

\bigskip

\centerline{
\begin{tabular}{|c|c|c|c|c|c|c|}
\hline\hline
$M$ &  &  & $\rho_{z}$ & $P$ & $C$ & $\Sigma_a$ \\ \hline\hline
$I_{\Pi }\otimes {\bf n}_{{\bf 1}}{\bf .\sigma }$ & ${\bf n_1}=
\left({\bf \hat{\imath}},{\bf \hat{\bf\jmath}}\right)$
& $\zeta=\frac{\pi }{2}$ & $-1$ & $+1$ & $+1$ &
$-1$ \\ \hline
& ${\bf n_1}={\bf \hat{k}}$ & $\zeta=0$ & $+1$ & $+1$ & $+1$ & $
-1 $ \\ \hline
$\Pi _{z}\otimes {\bf n}_{{\bf 2}}{\bf .\sigma }$ & ${\bf n_2}=
\left({\bf \hat{\imath}},{\bf \hat{\bf\jmath}}\right)$
& $\zeta=0$ & $+1$ & $-1$ & $-1$ & $-1$ \\ 
\hline
& ${\bf n_2}={\bf \hat{k}}$ & $\zeta=\frac{\pi }{2}$ & $-1$ & $-1$
& $-1$ & $-1$ \\ \hline
${\bf \nu }_{{\bf 1}}{\bf .\Pi }\otimes I_{\sigma }$ & ${\bf \nu_1}
={\bf \hat{\imath}}$ & $\xi=0$ & $+1$ & $+1$ & $+1$ & $+1$ \\ \hline
& ${\bf \nu_1} ={\bf \hat{\jmath}}$ & $\xi=\frac{\pi }{2}$ & $
-1 $ & $-1$ & $+1$ & $+1$ \\ \hline
${\bf \nu }_{{\bf 2}}{\bf .\Pi }\otimes {\bf n}_{{\bf B}}{\bf .\sigma }$ & $
{\bf \nu_2}= - {\bf \hat{\imath}}$ & $\xi=\frac{\pi }{2}$ & $-1$
& $+1$ & $-1$ & $+1$ \\ \hline
& ${\bf \nu_2}= {\bf\hat{\jmath}}$ & $\xi=0$ & $+1$ & $-1$ & $
-1$ & $+1$ \\ \hline\hline
\end{tabular}}

\bigskip

\centerline{Table 1: Discrete symmetries of the boundary conditions}

\bigskip

As representative examples, we consider below the boundary
conditions $M_1$ (diagonal) and $M_2$ (off-diagonal) separately.
We will calculate the form of the standing waves and the energy spectrum
for a nanotube with the same type of boundary condition on the right
(at $y = + L/2$) and on the left (at $y = - L/2$).
We introduce an index $u = \{R,L\} \equiv \pm 1$ to label
the right and left hand side so that the normal to the boundary,
defined with respect to the graphite coordinate system, is
${\bf n}_{{\bf B}}=u(\sin \eta ,\cos \eta ,0)$,
and we take into account the possibility of different mixing angles,
$\Lambda_u$,$\Upsilon_u$,$\Theta_u$ and $\Omega_u$, and vectors
${\bf n_1}=\left( u\cos\eta\sin\zeta_{u},-u\sin\eta\sin\zeta_{u},
\cos\zeta_{u}\right)$,
${\bf n_2}=\left( u\cos\eta\cos\zeta_{u},-u\sin\eta\cos\zeta_{u},
-\sin\zeta_{u}\right)$,
${\bf \nu_1}=\left( \cos\xi_{u},\sin\xi_{u},0\right)$
and ${\bf \nu_2}=\left( -\sin\xi_{u},\cos\xi_{u},0\right)$.

\subsection{Diagonal boundary conditions}

With the above definitions of the mixing angles,
the boundary condition $\Psi =M_{1}\Psi$ produces the following relations
between the components of the wavefunction at the interface:
\begin{eqnarray}
u\sin \left( \zeta _{u}+\Lambda _{u}\right) e^{-i\eta }\psi _{AK}-\left[
1+\cos \left( \zeta _{u}+\Lambda _{u}\right) \right] \psi _{BK} &=&0, \\
u\sin \left( \zeta _{u}-\Lambda _{u}\right) e^{+i\eta }\psi _{A\widetilde{K}
}-\left[ 1-\cos \left( \zeta _{u}-\Lambda _{u}\right) \right]
\psi_{B\widetilde{K}} &=&0.
\end{eqnarray}

The equations are diagonal in K-point space so do not describe intervalley
scattering. With these boundary conditions on the right (at $y = + L/2$) and
on the left (at $y = - L/2$), standing waves $\Psi_{1}$ corresponding to
K-point $K$ are created from combining $\Psi_{K}^{(R)}$ and $\Psi _{K}^{(L)}$
and are labelled by pseudo-helicity $\lambda = + s$,
and standing waves $\Psi_{2}$ corresponding to K-point $\widetilde{K}$ are
created from $\Psi _{\widetilde{K}}^{(R)}$ and $\Psi _{\widetilde{K}}^{(L)}$
and have label $\lambda = - s$:
\begin{eqnarray}
\Psi_{1} &=& {\cal N} \left(
\begin{array}{c}
e^{iq_1 y} + (-1)^{p_1} e^{is\zeta_m + is\Lambda_m - iq_1 y} \\ 
ise^{-i\eta }
\left[ e^{iq_1 y} - (-1)^{p_1} e^{is\zeta_m + is\Lambda_m - iq_1 y}\right] \\ 
0 \\ 
0
\end{array}
\right) ,  \\ 
\Psi_{2} &=& {\cal N} \left(
\begin{array}{c}
0 \\
0 \\ 
e^{iq_2 y} + (-1)^{p_2} e^{-is\zeta_m + is\Lambda_m - iq_2 y} \\ 
-ise^{-i\eta }
\left[ e^{iq_2 y} - (-1)^{p_2} e^{-is\zeta_m + is\Lambda_m - iq_2 y}\right]
\end{array}
\right) , 
\end{eqnarray}
where the normalisation factor is ${\cal N} = 1/(2\sqrt{L_c L})$
and the wavevectors are
\begin{eqnarray}
q_1 &=& \left( - s \zeta_p - s \Lambda_p + \pi p_{1} \right)/L , \\
q_2 &=& \left( + s \zeta_p - s \Lambda_p + \pi p_{2} \right)/L .
\end{eqnarray}
Here $\left\{ p_{1},p_{2}\right\} $ are integers such that $q_{1(2)} \geq 0$,
$\zeta_p = (\zeta_R + \zeta_L)/2$,
$\zeta_m = (\zeta_R - \zeta_L)/2$,
$\Lambda_p = (\Lambda_R + \Lambda_L)/2$, and
$\Lambda_m = (\Lambda_R - \Lambda_L)/2$.
Using $E=svq$ shows that the mixing angles $\zeta _{R}$ and $\zeta _{L}$ break
K-point degeneracy whereas
$\Lambda_{R}$ and $\Lambda _{L}$ break electron-hole symmetry.

\begin{figure}[t]
\centerline{\epsfxsize=0.4\hsize
\epsffile{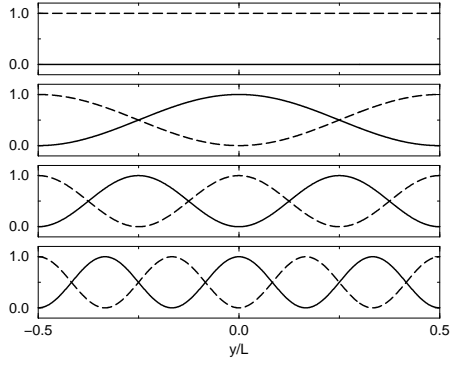}}
\caption{\label{fig:wfplot}
Plot of the modulus squared wavefunction $\left|\Psi_{1}\right|^2$ of
the lowest states for diagonal boundary conditions:
the solid lines show the second component $\left|\psi_{BK}\right|^2$ and
the dashed lines show the first component $\left|\psi_{AK}\right|^2$.
The lowest states $p_1=0,1,2,3$ are shown from top to bottom.
Values of the mixing angles are taken to be $\zeta_m = \Lambda_m = 0$
so that the boundary conditions are satisfied by
$\psi_{BK} =0$ at the ends of the nanotube $y = \pm L/2$.
}
\end{figure}

In order to understand the form of the wavefunctions, we set all mixing angles
equal to zero $\zeta_p = \zeta_m = \Lambda_p = \Lambda_m = 0$.
In this case the boundary conditions simplify to
$\psi_{BK} = \psi_{B\widetilde{K}} = 0$ at both ends of the nanotube,
and the components $\psi_{BK}$ and $\psi_{B\widetilde{K}}$ have the form of
standing wave solutions of the Schr\"odinger equation for a confined particle,
namely successive cosine and sine functions.
The component $\psi_{BK}$ is shown explicitly in Fig.~\ref{fig:wfplot}
(solid lines)
and the component $\psi_{AK}$,
which is proportional to the derivative of $\psi_{BK}$, is
shown by dashed lines.

%
%

\subsection{Off-diagonal boundary conditions}\label{ODBC}

The boundary condition $\Psi =M_{2}\Psi $ is equivalent to the following
relations between the components of the envelope wavefunction at the interface:
\begin{eqnarray}
\psi _{AK}+u\sin \Upsilon _{u}e^{+i\eta -i\xi _{u}}\psi _{A\widetilde{K}
}-\cos \Upsilon _{u}e^{-i\xi _{u}}\psi _{B\widetilde{K}} &=&0, \\
\psi _{BK}-u\sin \Upsilon _{u}e^{-i\eta -i\xi _{u}}\psi _{B\widetilde{K}
}-\cos \Upsilon _{u}e^{-i\xi _{u}}\psi _{A\widetilde{K}} &=&0.
\end{eqnarray}
The equations are off-diagonal in K space so describe intervalley scattering.
We label the standing waves as $\Psi_{1}$ with pseudospin eigenvalue
$\Sigma =+s$, created from combining $\Psi_{K}^{(R)}$ and
$\Psi _{\widetilde{K}}^{(L)}$, and $\Psi_{2}$ with pseudospin
eigenvalue $\Sigma =-s$,
created from combining $\Psi _{\widetilde{K}}^{(R)}$ and $\Psi_{K}^{(L)}$.
We find that
\begin{eqnarray}
\Psi_{1} &=&  {\cal N} \left(
\begin{array}{c}
e^{iq_1 y} \\ 
ise^{-i\eta + iq_1 y} \\ 
(-1)^{p_1} e^{is\Upsilon_m + i\xi_p -iq_1 y} \\ 
is (-1)^{p_1} e^{-i\eta + is\Upsilon_m + i\xi_p -iq_1 y}
\end{array}
\right) , \label{ow1} \\ 
\Psi_{2} &=& {\cal N} \left(
\begin{array}{c}
(-1)^{p_2} e^{is\Upsilon_m - i\xi_p - iq_2 y} \\ 
-is (-1)^{p_2} e^{-i\eta + is\Upsilon_m - i\xi_p -iq_2 y} \\ 
e^{iq_2 y}  \\ 
-ise^{-i\eta + iq_2 y}
\end{array}
\right) , \label{ow2}
\end{eqnarray}
where the normalisation factor is ${\cal N} = 1/(2\sqrt{L_c L})$
and the wavevectors are
\begin{eqnarray}
q_1 &=& \left( - s \Upsilon_p - \xi_m + \pi p_{1} \right) /L , \label{ODq1} \\
q_2 &=& \left( - s \Upsilon_p + \xi_m + \pi p_{2} \right) /L . \label{ODq2}
\end{eqnarray}
Here $\left\{ p_{1},p_{2}\right\} $ are integers such that $q_{1(2)}\geq 0$,
$\Upsilon_p = (\Upsilon_R + \Upsilon_L)/2$,
$\Upsilon_m = (\Upsilon_R - \Upsilon_L)/2$,
$\xi_p = (\xi_R + \xi_L)/2$, and
$\xi_m = (\xi_R - \xi_L)/2$.
The angle $\xi_m$ breaks degeneracy whereas $\Upsilon_p$
breaks electron-hole symmetry. 

%
\begin{figure}[t]
\centerline{\epsfxsize=0.4\hsize
\epsffile{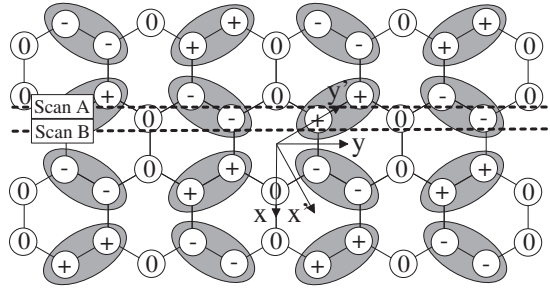}}
\vspace{0.3cm}
\caption{\label{fig:scans}
Relative amplitude of the wavefunction $\Psi_{{\rm tot},1} ({\bf r})\propto
\cos \left( {\bf K}.{\bf r} \mp \pi /6 \right)$ determined on the atomic sites,
following Fig.~1(d) in Ref.~\cite{rubio99}.
Dashed lines, labelled `scan A' and `scan B', are parallel to the
tube axis.
}
\end{figure}
%
%

The physical relevance of the envelope wavefunctions may be understood
by examining the total wavefunction $\Psi_{\rm tot}$, Eq.~(\ref{totalwf}),
that is constructed from linear combinations of products of envelope
wavefunctions with Bloch functions that vary rapidly in space on the atomic
scale.
If we only take into account the first term in the gradient expansion,
Eq.~(\ref{totalwf}), and the contribution from a single atomic orbital at 
${\bf r} = {\bf R_A}$ or ${\bf R_B}$, then $\Psi_{{\rm tot},1(2)}$
is the sum of two components of $\Psi_{1(2)}$, each multiplied by an
additional oscillating factor $\exp ( i{\bf K_{\mu }.R}_{{\bf j}} )$.
For example, if we set $\Upsilon_m = \xi_p = 0$ for an armchair tube
$\eta = \pi /6$ then
\begin{eqnarray}
\Psi_{{\rm tot},1} ( {\bf r} ) &\propto&
\left\{
\begin{array}{r@{\quad :\quad}l}
\cos \left( q_1 y + {\bf K}.{\bf r} \mp \pi /6 \right) & s(-1)^{p_1} = +1 \\
\sin \left( q_1 y + {\bf K}.{\bf r} \mp \pi /6 \right) & s(-1)^{p_1} = -1 
\end{array} \right. \nonumber \\
\Psi_{{\rm tot},2} ( {\bf r} ) &\propto&
\left\{
\begin{array}{r@{\quad :\quad}l}
\sin \left( q_2 y - {\bf K}.{\bf r} \pm \pi /6 \right) & s(-1)^{p_2} = +1 \\
\cos \left( q_2 y - {\bf K}.{\bf r} \pm \pi /6 \right) & s(-1)^{p_2} = -1 
\end{array} \right. \nonumber
\end{eqnarray}
where the upper sign refers to ${\bf r} = {\bf R_A}$
and the lower to ${\bf r} = {\bf R_B}$.
These equations reproduce the atomic scale variation of standing wave
patterns obtained by Rubio {\em et al} \cite{rubio99} with an additional
modulation due to the wavevector $q_{1(2)}$.
Fig.~\ref{fig:scans} is a schematic of the wavefunction amplitude
$\Psi_{{\rm tot},1} \propto
\cos \left( {\bf K}.{\bf r} \mp \pi /6 \right)$ determined on the atomic sites,
following Fig.~1(d) in Ref.~\cite{rubio99}.
Figs.~\ref{fig:linea} and~\ref{fig:lineb} show plots of
the modulus squared wavefunction for the four lowest states above
the Fermi level, evaluated along two different lines
parallel to the tube axis, labelled `scan A' and `scan B' in
Fig.~\ref{fig:scans}, respectively.
Fig.~\ref{fig:linea}, scan A, is for a line through the middle of
bonds making an angle with the tube axis and it tends to show a pair
of equidistant peaks within every Fermi wavelength whereas
Fig.~\ref{fig:lineb}, scan B, is for a line through bonds perpendicular
to the tube axis and it tends to show peak-pairing \cite{rubio99}.
In order to ensure that the successive wavefunctions are not degenerate,
we take $\xi_m = \pi /4$ and $\Upsilon_p = - \pi /2$ so that
the four lowest states above the Fermi level have energies
$E = \pi v/(4L)$, $3 \pi v/(4L)$, $5 \pi v/(4L)$, $7 \pi v/(4L)$ with
wavevector indices $p_1 = 0$, $p_2 = 0$, $p_1 = 1$, $p_2 = 1$,
and respective correspondence to the wavefunctions drawn schematically in
Fig.~1 (d), (a), (c), (b) of Ref.~\cite{rubio99}.
As well as a different long range modulation, due to different
values of $q_{1(2)}$, the successive wavefunctions show a distinct  
even/odd variation due to the different forms of pseudospin
eigenvectors $\Psi_{1}$ and $\Psi_{2}$.

%
\begin{figure}[t]
\centerline{\epsfxsize=0.4\hsize
\epsffile{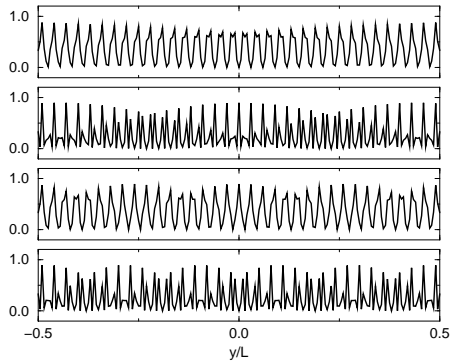}}
\caption{\label{fig:linea}
Plot of the modulus squared total wavefunction
$\left|\Psi_{tot}\right|^2$ (arbitrary units)
for off-diagonal boundary conditions that break valley degeneracy.
The wavefunction is evaluated along line A parallel to the axis of an
armchair nanotube $\eta = \pi/6$, length $L=50a$.
The four lowest energy states above the Fermi level are shown from top to
bottom.
Parameter values are $s=1$, $\zeta_m = \pi /4$, $\Upsilon_p = - \pi /2$, and
$\Upsilon_m = \zeta_p = 0$.
}
\end{figure}
%
%

%
\begin{figure}[t]
\centerline{\epsfxsize=0.4\hsize
\epsffile{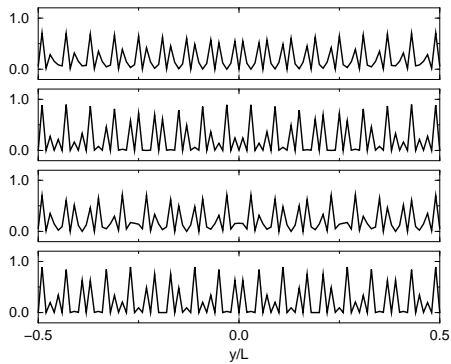}}
\caption{\label{fig:lineb}
Plot of the modulus squared total wavefunction
$\left|\Psi_{tot}\right|^2$ (arbitrary units)
for off-diagonal boundary conditions that break valley degeneracy.
The wavefunction is evaluated along line B parallel to the axis of an
armchair nanotube $\eta = \pi/6$, length $L=50a$.
The four lowest energy states above the Fermi level are shown from top to
bottom.
Parameter values are $s=1$, $\zeta_m = \pi /4$, $\Upsilon_p = - \pi /2$, and
$\Upsilon_m = \zeta_p = 0$.
}
\end{figure}
%

\section{Degenerate perturbation theory in the presence of an impurity}
\label{S:dpt}

In this section we derive $4 \times 4$ Hamiltonians of the
effective mass model describing a short range potential at different
positions ${\bf R}$ in the nanotube wall. Each effective Hamiltonian has
a different structure, depending on the location of the potential with
respect to the hexagonal graphite unit cell. The following subsections
describe different impurity positions as shown in Fig.~\ref{fig:positions}
and summarised in Table~2.

\bigskip 
\centerline{\begin{tabular}{||c||c||c||}
\hline\hline
Impurity position with respect & Label in & Discussed in \\ 
to the graphite unit cell & Fig.1 & subsection: \\ \hline\hline
atomic site & $A_{i},B_{i}$ & \ref{sub:A} \\ \hline
centre of unit cell & $C$ & \ref{sub:C} \\ \hline
half-way along a bond & $D_{i}$ & \ref{sub:D} \\ \hline\hline
\end{tabular}}
\bigskip

\centerline{Table 2: Impurity positions discussed in the following subsections}

\bigskip

As well as degeneracy arising from the real spin of electrons, it was
shown in the previous section that the energy levels of a clean metallic
nanotube may be degenerate due to pseudospin symmetry, depending on the
symmetry of the boundary conditions.
Now we would like to concentrate on the role of an additional perturbing
potential, so we will consider the case of degenerate levels in the clean
nanotube, and use degenerate perturbation theory to calculate the
level splitting due to the presence of the potential Hamiltonian.
The perturbation theory takes into account the interaction of the potential
with the degenerate levels, but neglects the effect of higher levels, so it is
valid for energy level shifts that are smaller than the spacing
$\Delta E=\pi v /L$ between pairs of unperturbed levels.
As before, we suppose that the $x$ axis is perpendicular
to the tube axis and we consider only the zero momentum transverse mode so
that $|E| < 2 \pi v /L_c$ where $L_c = |{\bf C_h}|$ is the circumference.
We will use off-diagonal boundary conditions M2 because they correspond to
the usual situation in metallic nanotubes \cite{m+f04}, so the unperturbed
degenerate wavefunctions are $\Psi_{1}$ and $\Psi_{2}$,
Eqs.~(\ref{ow1}) and (\ref{ow2}), respectively, with $\xi_R = \xi_L =0$
corresponding to pseudospin symmetry preserving boundaries,
$q_1 = q_2 \equiv q$, $p_1 = p_2 \equiv p$, and
$q = ( \pi p - s \Upsilon_p )/L$.

As explained in Section~\ref{S:emm}, we perform a gradient expansion
of the total wavefunction, Eq.~(\ref{totalwf}), and keep the lowest
order term.
Then, we calculate matrix elements
$V_{mn} = \int d{\bf r} \Psi_m^{\ast} \delta H \Psi_n$
between the clean wavefunctions
Eqs.~(\ref{ow1}) and (\ref{ow2}) and the effective Hamiltonians
in order to apply degenerate perturbation theory.
The matrix elements for a general effective Hamiltonian with
arbitrary coefficients are given in Appendix~A:
a particular position of the
potential will define the values of the arbitrary coefficients.
The positions of the potential we consider are shown with relation to the
hexagonal graphite unit cell in Fig.~\ref{fig:positions}.
They are near an A type atomic site, labelled A in the figure,
near a B type atomic site, labelled B,
near the centre of the unit cell, labelled C,
or near the half-way point between neighbouring atoms, labelled D.
Furthermore, we introduce a small additional deviation of the potential
position ${\bf \delta R}$, the orientation of which is shown in the figure
for the potential near the unit cell centre.
The labels $(x^{\prime},y^{\prime})$ represent the coordinate axes of the
graphite sheet, whereas labels $(x,y)$ represent the coordinate axes of
the nanotube, rotated by the chiral angle $\eta$.
The nanotube axis is parallel to the $y$ direction, and the direction
of the deviation of the potential position is described
by angle $\chi$ in the nanotube coordinates
${\bf \delta R} = (\delta R \cos\chi , \delta R \sin\chi)
\equiv (\delta X,\delta Y)$.

\subsection{Potential near an atomic site}\label{sub:A}

The origin of real space coordinates is placed in the centre of the
Wigner-Seitz unit cell and the perturbative potential is placed at
position ${\bf R} = {\bf R_0} + {\bf \delta R}$ near an arbitrary atomic site.
For example it may be near an A site, Fig.~\ref{fig:positions}, so that
${\bf R_0} = {\bf R}_{A}$ represents the exact position of the A atom,
and ${\bf \delta R}$ is a small additional deviation from it.
In deriving the effective mass model Hamiltonian, we take into account
nearest neighbour interactions: within nearest neighbour
distance $d=a/\sqrt{3}$ of the perturbative potential, there is one
A atom and three B atoms.
In addition to the gradient expansion, we perform an expansion in
the small additional deviation of the potential position ${\bf \delta R}$
in order to generate a number of effective Hamiltonians with
different symmetries.

The effective Hamiltonian $\delta H$ is a $4 \times 4$
Hamiltonian with matrix elements involving the Bloch function $\Phi _{m}^{(0)}$,
Eq.~(\ref{phi0}), and a short ranged potential
$\delta {\cal H} \left( {\bf r}\right)$ of strength $U$:
\begin{eqnarray}
\delta H_{nm} = \int d^{d}r \Phi_{n}^{(0)\ast}\left( {\bf r}\right)
\delta {\cal H} \left( {\bf r}\right) \Phi_{m}^{(0)}\left( {\bf r}\right) 
\end{eqnarray}
Integration with respect to fast degrees of freedom that vary on the
scale of the unit cell produces a product of Bloch functions $\Phi _{m}^{(0)}$
evaluated at the potential position and a delta function representing
the fact that the envelope wavefunctions interact with a localised
potential of scale less than the graphite lattice constant $a$:
\begin{eqnarray}
\delta H_{nm} \equiv
v_{a} U L^d \delta \left( {\bf r}-{\bf R} \right)
\Phi _{n}^{(0)\ast}\left( {\bf R}\right)
\Phi _{m}^{(0)}\left( {\bf R}\right) 
\end{eqnarray}
Here $v_a$ is the volume of the graphite unit cell.
There is a strong dependence of the phase factors contained within the
Bloch functions $\Phi _{m}^{(0)}$ on the position of the potential within
the graphite unit cell.
The Bloch functions also depend on $\pi$-type
atomic orbitals $\varphi_{j}$ on the non-equivalent atomic
sites $j = \left\{ A,B\right\}$ in the unit cell.
Since we consider the perturbative potential to be in the same plane as the
carbon atoms, we only need to describe the behaviour of the atomic orbitals
in the $(x,y)$ coordinates.
They are radially symmetric in the plane and for simplicity we model them as
$\varphi_{A/B}({\bf r}) \equiv \varphi ({\bf r})
= \varphi_{0} \exp (-|{\bf r}|/\lambda)$ where $\lambda \sim a/\sqrt{3}$.

\begin{figure}[t]
\centerline{\epsfxsize=0.4\hsize
\epsffile{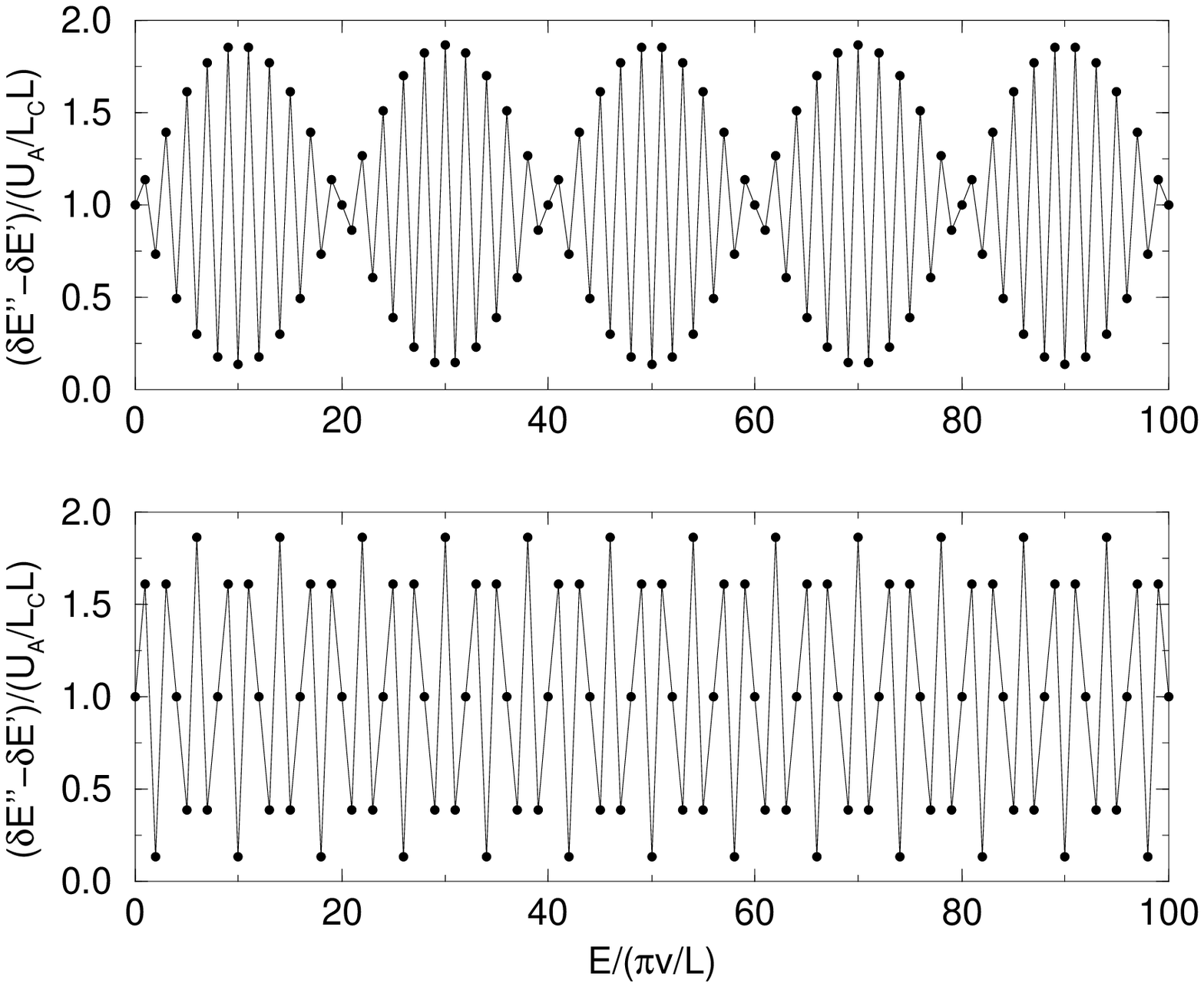}}
\caption{\label{fig:A00shift}
Splitting $\delta E^{\prime\prime} - \delta E^{\prime}$ of the
pairs of degenerate energy levels of a clean nanotube
Eq.~(\ref{A00shifts}) due to the effective Hamiltonian
$\delta H_{A}$ of a perturbative potential on an
A atomic site.
The symbols show the energy shift as a function of the energy of the
unperturbed levels, solid lines are a guide for the eye.
The upper curve is for the potential at $Y_0 = 0.025L$
(potential is one twentieth of the way from the centre of the nanotube to the end),
lower curve is for $Y_0 = 0.125L$
(potential is a quarter of the way from the centre of the nanotube to the end).
$U_{A} = v_{a}^{2}\varphi^{2}(0)U$ and parameter values are $s=1$,
$\kappa = 2\pi /3$, $\eta = \pi /6$, and $\Upsilon_p = \Upsilon_m = 0$.
}
\end{figure}

For the potential exactly on an A site, ${\bf \delta R} = 0$,
the effective Hamiltonian is
\begin{eqnarray}
\delta H_{A} =
v_{a}^{2}\varphi^{2}(0)U\delta \left( {\bf r} - {\bf R}\right) \left( 
\begin{array}{cccc}
1 & 0 & 0 & e^{-i\kappa } \\ 
0 & 0 & 0 & 0 \\ 
0 & 0 & 0 & 0 \\ 
e^{+i\kappa } & 0 & 0 & 1
\end{array}
\right) ,  \label{ha}
\end{eqnarray}
where $\kappa$
is a phase factor associated with intervalley scattering at the impurity
$\kappa = {\bf R_0}.({\bf K} - {\bf \widetilde{K}})$.
As expected for a potential on an atomic site,
this Hamiltonian breaks pseudospin-flip symmetry
$\delta H_{A} \neq
\rho_{z}^{-1} \delta H_{A}\rho_{z}$.
For completeness, we note that the equivalent effective Hamiltonian for
an impurity near a B site, Fig.~\ref{fig:positions}, is
\begin{eqnarray}
\delta H_{B}= v_{a}^{2}\varphi^{2}(0)U
\delta \left( {\bf r} - {\bf R} \right)
\left( 
\begin{array}{cccc}
0 & 0 & 0 & 0 \\ 
0 & 1 & e^{-i\kappa } & 0 \\ 
0 & e^{+i\kappa } & 1 & 0 \\ 
0 & 0 & 0 & 0
\end{array}
\right) . \label{hb}
\end{eqnarray}
Applying the general results for matrix elements given in Appendix~A to
the effective Hamiltonians $\delta H_{A/B}$,
we find that $V_{12}V_{21} = V_{11}V_{22}$ so that the energy level shifts
are $\delta E^{\prime} = 0$ and $\delta E^{\prime\prime} = V_{11} + V_{22}$.
In terms of the model parameters,
\begin{eqnarray}
\delta E^{\prime\prime} &=& \frac{v_{a}^{2}\varphi^{2}(0)U}{L_{c}L} 
\left[ 1 + a s (-1)^p \cos \left( \kappa + a \eta \right) 
\sin \left( 2 q Y_0 - s \Upsilon_m \right) \right] , 
\label{A00shifts} 
\end{eqnarray}
where $q = ( \pi p - s \Upsilon_p )/L$ and $-L/2 < Y_0 < L/2$ is the coordinate
of the perturbative potential along the nanotube axis.
Here the factor $a = \pm 1$ is used to distinguish between the case of
the potential being near an A site $a = 1$ or near a B site $a = -1$. 
There is an oscillating dependence of the energy level shift
on the index $p$ of the clean energy levels that has a period equal to
$1 / (Y_0/L)$.
In terms of energy, and the spacing of pairs of degenerate levels
$\Delta E = \pi v/L$, the period is
$\Delta E /(Y_0 /L) = \pi v /Y_0$.
Fig.~\ref{fig:A00shift} shows the splitting of the two levels as a function of
the energy for two different potential positions.
The upper curve is for $Y_0 = 0.025L$ (potential is one twentieth of the
way from the centre of the nanotube to the end), and shows an
oscillating pattern with period $40$, whereas 
the lower curve is for $Y_0 = 0.125L$ (potential is a quarter of the way
from the centre of the nanotube to the end), and shows an
oscillating pattern with period $8$.

%
%

\begin{figure}[t]
\centerline{\epsfxsize=0.4\hsize
\epsffile{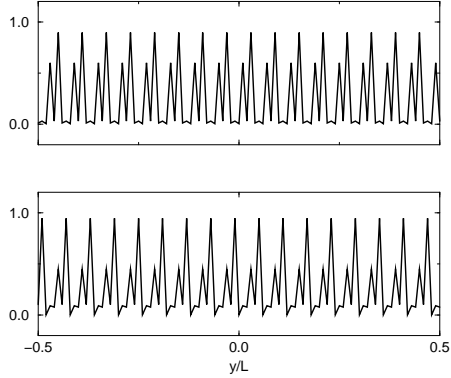}}
\caption{\label{fig:impb}
Plot of the modulus squared total wavefunctions
$\left|\Psi_{tot}\right|^2$ (arbitrary units) at the Fermi level ($q=0$)
in the presence of an impurity on an atomic site,
evaluated using degenerate perturbation theory.
The wavefunctions are evaluated along line B parallel to the axis of an
armchair nanotube $\eta = \pi/6$, length $L=50a$.
The standing wave corresponding to $\delta E = 0$ is shown on top,
that corresponding to $\delta E = V_{11} + V_{22}$ is below.
Parameter values are $s=1$, $p=0$, and
$\Upsilon_p = \Upsilon_m = \zeta_p = \zeta_m = 0$.
}
\end{figure}

The degenerate perturbation theory produces two new zero-order wavefunctions
that are linear combinations of the original ones.
We use them to plot the corresponding modulus squared total wavefunctions
$\left|\Psi_{tot}\right|^2$ near the Fermi level in Fig.~\ref{fig:impb}.
The special case of $q=0$ is considered, where the long-range variation
due to the envelope function is absent.
The top panel shows the wavefunction corresponding to $\delta E = 0$,
${\Psi}_1^{\prime} \propto \Psi_1 - (V_{11}/V_{12})\Psi_2$, that
has a matrix element with the effective Hamiltonian equal to zero
${V}_{11}^{\prime}  =
\int d{\bf r} {\Psi}_1^{\prime \ast}  \delta H {\Psi}_1^{\prime}  = 0$.
This wavefunction is zero on every third A site having the same phase factor
$\kappa$ as the impurity site.
The lower panel in Fig.~\ref{fig:impb} shows the wavefunction corresponding to
$\delta E = V_{11} + V_{22}$,
${\Psi}_2^{\prime} \propto \Psi_1 + (V_{22}/V_{12})\Psi_2$ that has a
non-zero matrix element with the effective Hamiltonian.
It has a sharp peak on every third A site where the other standing wave
is zero.

%
%

The effective Hamiltonians Eqs.~(\ref{ha}) and (\ref{hb}) for a potential
exactly on an atomic site break axis reflection symmetry.
In order to demonstrate the role of symmetry, we take the sum of Hamiltonians
arising from potentials on adjacent A and B atoms with the same
component along the tube axis: for example, positions $A_1$ and $B_1$ in
Fig.~\ref{fig:positions}. In this case the Hamiltonian is
\begin{eqnarray}
&&\delta H_{A} + \delta H_{B} =
v_{a}^{2}\varphi^{2}(0)U\delta \left( {\bf r} - {\bf R}\right) \left( 
\begin{array}{cccc}
1 & 0 & 0 & e^{-i\kappa } \\ 
0 & 1 & e^{-i\beta } & 0 \\ 
0 & e^{+i\beta } & 1 & 0 \\ 
e^{+i\kappa } & 0 & 0 & 1
\end{array}
\right) ,  
\end{eqnarray}
where $\kappa = {\bf R_A}.({\bf K} - {\bf \widetilde{K}})$
and $\beta = {\bf R_B}.({\bf K} - {\bf \widetilde{K}})$.
We find that
\begin{eqnarray}
V_{12}V_{21} &\propto& 
\left[ \cos \left( \kappa + \eta \right) + \cos \left( \beta - \eta \right)
\right]^2 , \nonumber \\
\left( V_{11} - V_{22} \right)^2 &\propto&  
\left[ \sin \left( \kappa + \eta \right) - \sin \left( \beta - \eta \right)
\right]^2 . \nonumber
\end{eqnarray}
For the positions $A_1$ and $B_1$ in Fig.~\ref{fig:positions},
the phase factors are $\kappa = 2\pi /3$ and $\beta = 0$ in which
case the Hamiltonian
$\delta H_{A1} + \delta H_{B1}$
preserves axis reflection symmetry
$\Sigma_a^{-1} \delta H \Sigma_a = \delta H$ and $V_{12}V_{21} = 0$
for an armchair tube $\eta = \pi /6$.
There is no mixing of the pseudospin eigenfunctions,
but degeneracy is still broken $V_{11} - V_{22} \neq 0$.
Alternatively, using the unitary
transformation ${\cal U}$, Eq.~(\ref{trans}), to change to the system where
the clean wavefunctions are eigenvalues of $\sigma_z$, it is possible to
produce a matrix that has no off-diagonal spin parts and clearly does not
mix the pseudospin eigenfunctions:
\begin{eqnarray}
&&{\cal U}^{-1} \left( \delta H_{A1} + \delta H_{B1} \right) {\cal U} =
v_{a}^{2}\varphi^{2}(0)U\delta \left( {\bf r} - {\bf R}\right) \left( 
\begin{array}{cccc}
1 & 0 & e^{-\pi i/3 } & 0 \\ 
0 & 1 & 0 & e^{2\pi i/3 } \\ 
e^{\pi i/3 } & 0 & 1 & 0 \\ 
0 & e^{-2\pi i/3 } & 0 & 1
\end{array}
\right) .  
\end{eqnarray}
%

%
\begin{figure}[t]
\centerline{\epsfxsize=0.4\hsize
\epsffile{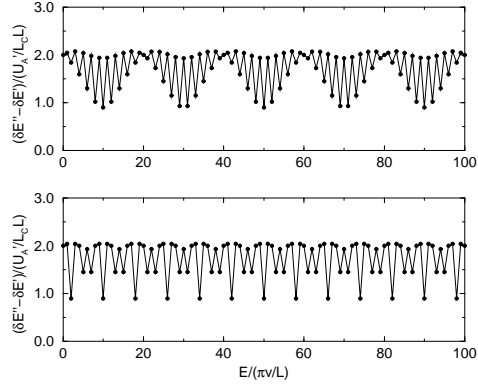}}
\caption{\label{fig:A01shift}
Splitting $\delta E^{\prime\prime} - \delta E^{\prime}$
of the pairs of degenerate energy levels of a clean nanotube
Eq.~(\ref{A01shifts}) due to the effective Hamiltonian
$\delta H_{A}^{\prime}$ of a perturbative potential with
a first order deviation ${\bf \delta R}$ from an A atomic site.
The symbols show the splitting as a function of the
energy of the unperturbed levels, lines are a guide for the eye.
The upper curve is for the potential at $Y_0 = 0.025L$
(potential is one twentieth of the way from the centre of the nanotube to the end),
lower curve is for $Y_0 = 0.125L$
(potential is a quarter of the way from the centre of the nanotube to the end).
Parameter values are $s=1$, $\kappa = 2\pi /3$, $\eta = \pi /6$,
$\Upsilon_p = \Upsilon_m = 0$, and the angle of deviation of the potential
is $\chi = \pi/4$.
}
\end{figure}
%

Returning to a single impurity potential on an atomic site, and taking into
account first order terms in the deviation ${\bf \delta R}$
of the potential position, we find an effective Hamiltonian with
a different structure:
\begin{eqnarray}
\delta H_{A/B}^{\prime} =
U_{A/B}^{\prime}\delta \left( {\bf r} - {\bf R}\right) \left( 
\begin{array}{cccc}
0 & iae^{-i\widetilde{\chi}} & -ie^{ia\widetilde{\chi}-i\kappa } & 0 \\ 
-iae^{+i\widetilde{\chi}} & 0 & 0 & -ie^{ia\widetilde{\chi}-i\kappa } \\ 
ie^{-ia\widetilde{\chi}+i\kappa } & 0 & 0 & iae^{-i\widetilde{\chi}} \\ 
0 & ie^{-ia\widetilde{\chi}+i\kappa } & -iae^{+i\widetilde{\chi}} & 0
\end{array}
\right) , 
\end{eqnarray}
where
$U_{A/B}^{\prime} = 3\left| {\bf \delta R}\right|
e^{-d/\lambda }v_{a}^{2}\varphi^{2}(\left| {\bf \delta R}\right| )U/2\lambda$
and $\widetilde{\chi} = \chi - \eta$ is the angle of the deviation
${\bf \delta R}$ in the graphite coordinates as shown in
Fig.~\ref{fig:positions}.
The factor $a = \pm 1$ is used to distinguish between the case of
the potential being near an A site $a = 1$ or near a B site $a = -1$.
We find that the energy level shifts are
\begin{eqnarray}
\delta E &=&
- \frac{U_{A/B}^{\prime}}{L_{c}L}
(-1)^p \cos \left( 2 q Y_0 - s \Upsilon_m \right) 
\sin \left( \kappa + a \eta - a \chi \right) \nonumber \\
&& \!\!\!\!\!\! \!\!\!\!\!\! \pm \, \frac{U_{A/B}^{\prime}}{L_{c}L}
\sqrt{\left[1 + a s (-1)^p \sin \left( 2 q Y_0 - s \Upsilon_m \right) 
\cos \left( \kappa + a\eta \right)\right]  
\left[1 +
a s (-1)^p \sin \left( 2 q Y_0 - s \Upsilon_m \right) 
\cos \left( \kappa + a\eta - 2\chi \right)\right]}  . 
\label{A01shifts}
\end{eqnarray}
where $\chi$ is the angle of the deviation of the potential in
the nanotube coordinates. 
Fig.~\ref{fig:A01shift} shows the splitting of the energy levels as a
function of the energy for two different potential positions.
The upper curve is for $Y_0 = 0.025L$ (potential is one twentieth of the
way from the centre of the nanotube to the end), and shows an
oscillating pattern with period $40$, whereas 
the lower curve is for $Y_0 = 0.125L$ (potential is a quarter of the way
from the centre of the nanotube to the end), and shows an
oscillating pattern with period $8$.
The oscillation periods are the same as for the Hamiltonian 
$\delta H_{A}$, but this time there is a shift of
both of the energy levels, one positive, one negative, instead of
one of the levels remaining stationary while the other moves.

\subsection{Impurity at the centre of the unit cell}\label{sub:C}

\begin{figure}[t]
\centerline{\epsfxsize=0.4\hsize
\epsffile{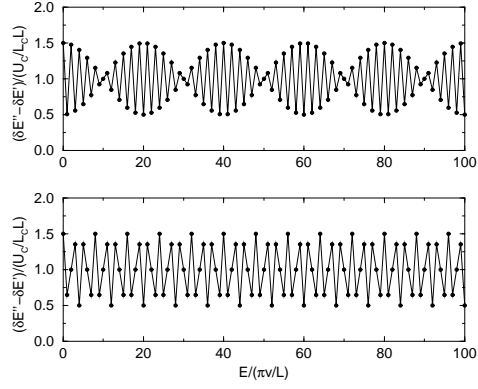}}
\caption{\label{fig:C02shift}
Splitting $\delta E^{\prime\prime} - \delta E^{\prime}$
of the pairs of degenerate energy levels of a clean nanotube
Eq.~(\ref{C02shifts}) due to the effective Hamiltonian
$\delta H_{C}$ of a perturbative potential near the
centre of the graphite unit cell.
The symbols show the energy shift as a function of the energy of the
unperturbed levels, solid lines are a guide for the eye.
The upper curve is for the potential at $Y_0 = 0.025L$
(potential is one twentieth of the way from the centre of the nanotube to the end),
lower curve is for $Y_0 = 0.125L$
(potential is a quarter of the way from the centre of the nanotube to the end).
Parameter values are $s=1$, $\kappa = 2\pi /3$, $\eta = \pi /6$,
$\Upsilon_p = \Upsilon_m = 0$, and the angle of deviation of the potential
is $\chi = \pi/4$.
}
\end{figure}

In this section, we consider the perturbative potential to be placed
near the centre of the graphite unit cell,
position C in Fig.~\ref{fig:positions}.
For the zeroth order gradient term, we find that the effective Hamiltonian
for the potential exactly at the centre of the unit cell is equal to zero:
such a position does not break the rotational symmetry of graphene.
The first non-zero contribution arises from a quadratic deviation
from the centre of the unit cell:
\begin{eqnarray}
\delta H_{C} =
iU_{C} \delta \left( {\bf r} - {\bf R} \right) \left( 
\begin{array}{cccc}
1 & e^{2i\widetilde{\chi}} &
-e^{-i\kappa } & -e^{2i\widetilde{\chi}-i\kappa } \\ 
e^{-2i\widetilde{\chi}} & 1 &
-e^{-2i\widetilde{\chi}-i\kappa } & -e^{-i\kappa} \\ 
-e^{+i\kappa } & -e^{2i\widetilde{\chi}+i\kappa } &
1 & e^{+2i\widetilde{\chi}} \\ 
-e^{-2i\widetilde{\chi}+i\kappa } & -e^{+i\kappa } &
e^{-2i\widetilde{\chi}} & 1
\end{array}
\right) , 
\end{eqnarray}
where $U_{C}=
\left( 3\left| {\bf \delta R}\right| / (2\lambda )\right)^{2}
v_{a}^{2}\varphi^{2}(d)U$.
Applying degenerate perturbation theory in the same way as before we find
that the energy level shifts are $\delta E^{\prime} = 0$ and
$\delta E^{\prime\prime} = V_{11} + V_{22}$.
In terms of the model parameters,
\begin{eqnarray}
\delta E^{\prime\prime} &=& \frac{2U_{C}}{L_{c}L}
\left[ 1 - (-1)^p \cos \left( 2 q Y_0 - s \Upsilon_m \right) 
\cos \kappa + \, \, s (-1)^p \sin \left( 2 q Y_0 - s \Upsilon_m \right)
\sin \left( 3\eta - 2\chi \right) 
\sin \kappa \right] , 
\label{C02shifts}
\end{eqnarray}
The results are similar to those for the potential exactly on an atomic site:
one of the energy levels does not move and corresponds to a linear combination
of clean wavefunctions that has zero matrix element with the effective
Hamiltonian, whereas the other energy level suffers a shift that oscillates
with the index $p$ and has a period equal to $1 / (Y_0/L)$.
Fig.~\ref{fig:C02shift} shows the energy level splitting as a function of
the energy for two different potential positions.
The upper curve is for $Y_0 = 0.025L$ (potential is one twentieth of the
way from the centre of the nanotube to the end), and shows an
oscillating pattern with period $40$, whereas 
the lower curve is for $Y_0 = 0.125L$ (potential is a quarter of the way
from the centre of the nanotube to the end), and shows an
oscillating pattern with period $8$.
The oscillation of the level splitting as a function of energy with a period
determined by the position $Y_0$ of the impurity along the tube axis,
$-L/2 < Y_0 < L/2$, may be understood as arising from 
the slow spatial modulation of the envelope wavefunctions since,
for standing waves, the positions of peaks and nodes vary as
a function of energy. Therefore the extent to which they scatter from a given
impurity position also depends on their energy.

%
%

The dependence of the energy level shift on $\eta$ and $\chi$ arises because
the position of the impurity determines the extent of degeneracy breaking.
As a special case, we note that when the angle of deviation of the
impurity is $\chi = \pi /2$ (see Fig.~\ref{fig:positions}), the Hamiltonian
$\delta H_{C}$ preserves axis reflection symmetry
for an armchair tube and the factor $\sin \left( 3\eta - 2\chi \right) = -1$.
Degeneracy is still broken but there are no 
off-diagonal matrix elements $V_{12} = V_{21} = 0$ and
no mixing of the pseudospin eigenvectors.
Alternatively, using the unitary
transformation ${\cal U}$, Eq.~(\ref{trans}), to change to the system where
the clean wavefunctions are eigenvalues of $\sigma_z$, it is possible to
produce a matrix that has no off-diagonal spin parts and clearly does not
mix the pseudospin eigenfunctions:
\begin{eqnarray}
{\cal U}^{-1} \delta H_{C} {\cal U}
= 2iU_{C} \delta \left( {\bf r} - {\bf R} \right)
\left( 
\begin{array}{cccc}
0 & 0 & 0 & 0 \\ 
0 & 1 & 0 & -e^{-i\kappa} \\ 
0 & 0 & 0 & 0 \\ 
0 & -e^{i\kappa} & 0 & 1
\end{array}
\right) .  
\end{eqnarray}
%

\subsection{Impurity half-way between atomic sites}\label{sub:D}

In this section, we consider the perturbative potential to be placed
near the half-way point between two neighbouring atoms,
position D in Fig.~\ref{fig:positions}.
There is a non-zero contribution of the zeroth order gradient term
for the potential exactly at the half-way point:
\begin{eqnarray}
\delta H_{D} =
iU_{D} \delta \left( {\bf r} - {\bf R} \right) \left( 
\begin{array}{cccc}
1 & e^{-i(\alpha - \beta )} &
e^{-i(\alpha + \beta )} & e^{-2i\alpha} \\ 
e^{i(\alpha - \beta )} & 1 &
e^{-2i\beta} & e^{-i(\alpha + \beta )} \\ 
e^{i(\alpha + \beta )} & e^{2i\beta} &
1 & e^{-i(\alpha - \beta )} \\ 
e^{2i\alpha} & e^{i(\alpha + \beta )} &
e^{i(\alpha - \beta )} & 1
\end{array}
\right) , 
\end{eqnarray}
where $U_{D} = v_{a}^{2} \varphi^{2}(d/2) U$,
and the phase factors $\alpha = {\bf K}.{\bf R_A}$ and
$\beta = {\bf K}.{\bf R_B}$ are evaluated for the two atomic positions
${\bf R_A}$ and ${\bf R_B}$ nearest the impurity.
We find that the matrix elements are
\begin{eqnarray}
V_{11/22} &=& \frac{U_{D}}{2L_{c}L}
\left\{ 2 \pm 2s \sin \left( \eta + \alpha - \beta \right) \right. \nonumber \\
&& + 2 (-1)^p \cos \left( \alpha + \beta \pm (2 q Y_0 - s \Upsilon_m) \right)
\nonumber \\
&&  \pm s (-1)^p 
\sin \left( 2\alpha + \eta \pm (2 q Y_0 - s \Upsilon_m) \right)  \nonumber \\
&&  \left. \mp s (-1)^p
\sin \left( 2\beta - \eta \pm (2 q Y_0 - s \Upsilon_m) \right) \right\}
, \nonumber \\
V_{12}V_{21} &=& \left( \frac{U_{D}}{2L_{c}L} \right)^2
\left\{  \cos \left( 2\alpha + \eta \right)
+ \cos \left( 2\beta - \eta \right)  \right. \nonumber \\
&& \left. + 2 (-1)^p \cos \left( \eta + \alpha - \beta \right)
\cos \left( 2 q Y_0 - s \Upsilon_m \right) \right\}^2 . \nonumber
\end{eqnarray}
Generally, there are two non-zero energy shifts, the exact values of
which depend on the phase factors $\alpha$ and $\beta$ that may take the
values $0$, $2\pi /3$, or $-2\pi /3$, depending on the particular
position: there are six D positions shown in
Fig.~\ref{fig:positions}. 

However, as a special case, we note that for positions $D_1$ and $D_4$ in
Fig.~\ref{fig:positions} the Hamiltonian
$\delta H_{D}$ preserves axis reflection symmetry
$\Sigma_a^{-1} \delta H_{D} \Sigma_a = \delta H_{D}$
for an armchair tube $\eta = \pi /6$.
Degeneracy is still broken but there are no 
off-diagonal matrix elements $V_{12} = V_{21} = 0$ and
no mixing of the pseudospin eigenvectors.
For example, $\alpha = 0$ and $\beta = 2\pi /3$ for position $D_1$
and, using the unitary transformation ${\cal U}$, Eq.~(\ref{trans}),
to change to the system where
the clean wavefunctions are eigenvalues of $\sigma_z$, it is possible to
produce a matrix that has no off-diagonal spin parts and clearly does not
mix the pseudospin eigenfunctions:
\begin{eqnarray}
{\cal U}^{-1} \delta H_{D} {\cal U}
= 2iU_{D} \delta \left( {\bf r} - {\bf R} \right)
\left( 
\begin{array}{cccc}
0 & 0 & 0 & 0 \\ 
0 & 1 & 0 & -e^{\pi i/3} \\ 
0 & 0 & 0 & 0 \\ 
0 & -e^{-\pi i/3} & 0 & 1
\end{array}
\right) .  
\end{eqnarray}
Since pseudospin is related to the underlying molecular orbital state
\cite{mceuen99}, the statement that impurities
preserving mirror reflection in the nanotube axis manage to
break degeneracy without mixing the pseudospin eigenvectors
is equivalent to saying that impurities
preserving mirror reflection do not result in 
mixing of the bonding $\pi$ and antibonding
$\pi^{\ast}$ energy bands \cite{del98,mat98,ando99,choi00,song02}.

\section{Conclusion}
%
In this paper, we considered degeneracy breaking due to short-ranged
impurities in finite, single-wall, metallic carbon nanotubes.
The effective mass model was used to describe the slowly varying spatial
envelope wavefunctions of spinless electrons near the Fermi level at two
inequivalent valleys (K-points) in terms of the four component Dirac equation
for massless fermions, with the role of spin assumed by pseudospin due to
the relative amplitude of the wave function on the sublattice atoms.
Using boundary conditions at the ends of the tube that neither
break valley degeneracy nor mix pseudospin eigenvectors,
we used degenerate perturbation theory to study the influence of impurities.
The position of a short-ranged impurity potential within the hexagonal graphite unit
cell produces a particular $4\times 4$ matrix structure of the
corresponding effective Hamiltonian, and the symmetry of the Hamiltonian
with respect to pseudospin flip and mirror reflection in the nanotube axis
is related to degeneracy breaking and pseudospin mixing, respectively.
Table 3 shows a summary of the position dependence for an armchair tube [axis is parallel to
the y-axis in Fig.~(\ref{fig:positions})].
It shows how the character of an impurity determines the extent of
valley degeneracy breaking, resulting in the possibility to observe experimentally either
twofold or fourfold periodicity of shell filling \cite{cob02}.
For example, an impurity on an atomic site will break valley degeneracy
and tend to give twofold periodicity, corresponding to spin degeneracy only, whereas a potential
at the centre of the graphite unit cell will not break valley degeneracy
and it will preserve fourfold periodicity corresponding to both spin and valley degeneracy.

\bigskip

\centerline{\begin{tabular}{||c||c||c||c||}
\hline\hline
Impurity position with respect & Label in  & Breaks valley  & Breaks axis 
\\ 
to the graphite unit cell & Fig.1 & degeneracy & reflection symmetry \\ 
\hline\hline
atomic site & $A_{i},B_{i}$ & yes & yes \\ \hline
centre of unit cell & $C$ & no & no \\ \hline
midway along a bond that is & $D_{1},D_{4}$ & yes & no \\ 
perpendicular to tube axis &  &  &  \\ \hline
midway along a bond & $D_{2},D_{3},D_{5},D_{6}$ & yes & yes \\ 
at $30%
{{}^\circ}%
$ angle with tube axis &  &  &  \\ \hline\hline
\end{tabular}}

\bigskip

\centerline{Table 3: The dependence of degeneracy breaking on the impurity position for an armchair nanotube}

\bigskip

In addition to position dependence on the scale of the
graphite unit cell, the level splitting displays a sinusoidally varying
energy dependence that has a period determined by
the position $Y_0$ of the impurity along the tube axis $-L/2 < Y_0 < L/2$.
This arises from 
the slow spatial modulation of the envelope wavefunctions since,
for standing waves, the location of peaks and nodes varies as
a function of energy. Therefore the extent to which they scatter from a given
impurity position also depends on their energy.
It means that, in experimental observations, the shell filling properties may not
be the same in different parts of the spectrum.

\acknowledgments
The authors thank EPSRC for financial support.

\begin{appendix}
\section{General form of the matrix elements of degenerate perturbation theory\label{A:gf}}

In this appendix, we present expressions for matrix elements
$V_{mn} = \int d{\bf r} \Psi_m^{\ast} \delta H \Psi_n$
between the clean wavefunctions Eqs.~(\ref{ow1}) and (\ref{ow2}) and a
general effective Hamiltonian with arbitrary coefficients.
We set $\xi_R = \xi_L =0$ corresponding to pseudospin symmetry preserving
boundaries, $q_1 = q_2 \equiv q$, $p_1 = p_2 \equiv p$, and
$q = ( \pi p - s \Upsilon_p )/L$.
The only constraints we apply to the general effective Hamiltonian are
due to hermicity and time reversal symmetry, because these constraints
are obeyed by every specific effective Hamiltonian that we derive. 
The results are used in Section~\ref{S:dpt} where a particular position of the
potential corresponds to particular values of the arbitrary coefficients.

We use the constraints of hermicity
and time reversal symmetry to write a general effective Hamiltonian as
\begin{eqnarray}
\delta H =
U \delta \left( {\bf r} - {\bf R}\right) \left( 
\begin{array}{cccc}
a & ce^{+i\delta } & me^{-i\mu } & l e^{-i\alpha } \\ 
ce^{-i\delta } & b & we^{-i\beta } & me^{-i\mu } \\ 
me^{i\mu } & we^{i\beta } & b & ce^{+i\delta } \\ 
l e^{i\alpha } & me^{i\mu } & ce^{-i\delta } & a
\end{array}
\right) ,  
\end{eqnarray}
where all the variables represent arbitrary real numbers.
We find that the matrix elements are
\begin{eqnarray}
V_{11/22} &=& \frac{U}{2L_{c}L}
\left\{ a + b \pm 2sc \sin \left( \eta - \delta \right) \right. \nonumber \\
&&\!\!\!  + 2m (-1)^p \cos \left( \mu \pm (2 q Y_0 - s \Upsilon_m) \right)
\nonumber \\
&&\!\!\!  \pm sl (-1)^p \sin \left( \alpha + \eta \pm (2 q Y_0 - s \Upsilon_m) \right)
\nonumber \\
&&\!\!\! \!\!\!  \left. \mp sw (-1)^p \sin
\left( \beta - \eta \pm (2 q Y_0 - s \Upsilon_m) \right)\right\} ,\label{gzd}\\
V_{12}V_{21} &=& \left( \frac{U}{2L_{c}L} \right)^2
\left\{  (-1)^p \left( a - b \right) \sin \left( 2 q Y_0 - s \Upsilon_m \right)
\right. \nonumber \\
&&\!\!\!  + 2sc (-1)^p \cos\left( \eta - \delta \right)
\cos\left( 2 q Y_0 - s \Upsilon_m \right) \nonumber \\
&&\!\!\! \!\!\!  \left. + sl \cos\left( \alpha + \eta \right)
+ sw \cos\left( \beta - \eta \right) \right\}^2 . \label{gzo}
\end{eqnarray}
where the upper sign in Eq.~(\ref{gzd}) refers to $V_{11}$
and the lower to $V_{22}$.

The Hamiltonian $\delta H$ preserves axis reflection symmetry
$\Sigma_a^{-1} \delta H \Sigma_a = \delta H$ for an armchair tube
$\eta = \pi /6$ if $b=a$, $w=l$, $\delta = 2\pi /3$ and
$\beta = \alpha - 2\pi /3$.
Degeneracy is still broken but there are no 
off-diagonal matrix elements $V_{12} = V_{21} = 0$ and
no mixing of the pseudospin eigenvectors.
Using the unitary transformation ${\cal U}$, Eq.~(\ref{trans}),
to change to the system where the clean wavefunctions are eigenvalues of
$\sigma_z$, it is possible to show explicitly that the Hamiltonian preserving
axis reflection symmetry has no off-diagonal
spin parts and clearly does not mix the pseudospin eigenfunctions:
\begin{eqnarray}
{\cal U}^{-1} \delta H {\cal U}=
U \, \delta \left( {\bf r} - {\bf R} \right) \left( 
\begin{array}{cccc}
a-c & 0 & \widetilde{m}^{\ast} + \widetilde{l}^{\ast} & 0 \\ 
0 & a+c & 0 & \widetilde{m}^{\ast} - \widetilde{l}^{\ast} \\ 
\widetilde{m} + \widetilde{l} & 0 & a-c & 0 \\ 
0 & \widetilde{m} - \widetilde{l} & 0 & a+c
\end{array}
\right) , 
\end{eqnarray}
where $\widetilde{m} = me^{i\mu }$ and
$\widetilde{l} = le^{i\alpha - \pi i/3}$.

The Hamiltonian $\delta H$ preserves pseudospin-flip symmetry
$\rho_{z}^{-1} \delta H \rho_{z} = \delta H$
if $b=a$, $c=0$, $\mu = 0$ (or $m=0$), and $l = -w$ and $\alpha = -\beta$
(or $l=w=0$), in which case $V_{12}V_{21} = V_{11} - V_{22} = 0$ meaning
that degeneracy is not broken.

\section{Non-perturbative determination of the spectrum in the presence
of an impurity\label{A:ndpt}}

In this section, we present a non-perturbative calculation of the energy
level spectrum in the presence of an impurity.
We consider the additional potential to be placed at an arbitrary position
$Y_0$ along the tube $-L/2 \leq Y_0 < L/2$, and we use the off-diagonal
boundary conditions, Section~\ref{ODBC}, at the ends of the tube $y = \pm L/2$.  
Since the potential is a delta function in space, the wavefunctions away from
it are simply the solutions of the clean Hamiltonian.
However, the delta function potential does introduce non-trivial matching
conditions at $Y_0$ for the standing waves to the left and the right.
In general, we have an equation of the form 
\begin{eqnarray}
\left[ -iv{\bf \alpha .\nabla } +
\delta \left( {\bf r} - {\bf R} \right) V \right] \Psi = E \Psi
\label{NP1}
\end{eqnarray}
where $V$ is a $4\times 4$ matrix as found in Section~\ref{S:dpt}.
To produce the matching conditions, we integrate the equation with
respect to $y$ over a vanishingly small interval
$Y_0 - \delta \leq y \leq Y_0+ \delta$ near the additional potential.
The first term in Eq.~(\ref{NP1}) gives a discontinuity in the components
$\psi_{m}$ of the envelope wavefunction at the potential position,
producing expressions such as
$\psi_{1} \left( Y_0+\delta \right) - \psi_{1} \left( Y_0-\delta \right)$.
The second term $\delta \left( {\bf r} - {\bf R} \right) V \Psi$
gives the value of the wavefunction components at the potential position and
the term on the right hand side of Eq.~(\ref{NP1}), $E\Psi$, gives zero
contribution: although $\Psi$ is not necessarily continuous, it is not infinite.
The wavefunctions are then determined, using the resulting matching conditions,
and the energy level spectrum is found.
As before, we will consider only metallic nanotubes
with arbitrary chiral angle $\eta$.
We suppose that the $x$ axis is perpendicular to the tube axis and we
consider only the zero momentum transverse mode so that
$|E| < 2 \pi v /L_c$ where $L_c = |{\bf C_h}|$ is the circumference.

%
\begin{figure}[t]
\centerline{\epsfxsize=0.4\hsize
\epsffile{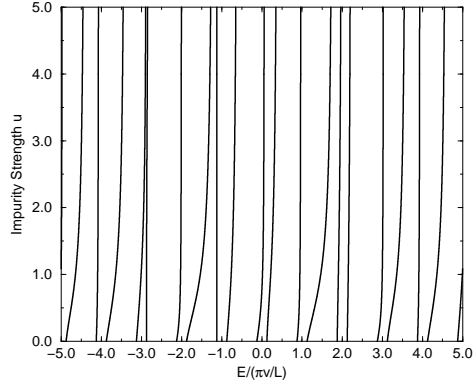}}
\caption{\label{fig:nd1}
Position of energy levels as a function of the strength of a potential
on an A atomic site in the presence of degeneracy breaking due to off-diagonal
boundary conditions $\xi_m = \pi /8$.
The position of the impurity is $Y_0 = 0.125L$
(potential is a quarter of the way from the centre of the nanotube to the end),
and other parameter values are $\kappa = 2\pi /3$, $\eta = \pi /6$,
$\xi_p = \Upsilon_p = \Upsilon_m = 0$.
}
\end{figure}
%

In order to demonstrate what is in principle possible, we consider in
detail the case of the additional potential exactly on an A site
with the following effective Hamiltonian:
\begin{eqnarray}
\delta H_{A} =
U_{A} \, \delta \left( {\bf r} - {\bf R}\right) \left( 
\begin{array}{cccc}
1 & 0 & 0 & e^{-i\kappa } \\ 
0 & 0 & 0 & 0 \\ 
0 & 0 & 0 & 0 \\ 
e^{+i\kappa } & 0 & 0 & 1
\end{array}
\right) ,  
\end{eqnarray}
where $\kappa$
is a phase factor associated with intervalley scattering at the impurity
$\kappa = {\bf R_0}.({\bf K} - {\bf \widetilde{K}})$.
It results in the following matching conditions at the impurity,
\begin{eqnarray}
\psi_{1} \left( Y_0+\delta \right) - \psi_{1} \left( Y_0-\delta \right)
&=& 0  , \\
\psi_{2} \left( Y_0+\delta \right) - \psi_{2} \left( Y_0-\delta \right)
&=& u e^{-i\eta}
\left[ \psi_{1} \left( Y_0 \right) + e^{-i\kappa} \psi_{4}\left( Y_0 \right)
\right] ,  \\
\psi_{3} \left( Y_0+\delta \right) - \psi_{3} \left( Y_0-\delta \right)
&=& u e^{i\eta} 
\left[ e^{i\kappa} \psi_{1}\left( Y_0 \right) + \psi_{4}\left( Y_0 \right)
\right] ,  \\
\psi_{4} \left( Y_0+\delta \right) - \psi_{4} \left( Y_0-\delta \right)
&=& 0  ,  
\end{eqnarray}
where $u = v_{a}^{2}\varphi^{2}(0)U/ v L_c$.
Using these matching conditions, we find that the wavevectors are given by
solutions of the following equation:
\begin{eqnarray}
&& 0 =  \sin \left( qL + s\Upsilon_p + \xi_m \right)
\sin \left( qL + s\Upsilon_p - \xi_m \right) \nonumber \\
&& \qquad - \frac{u}{2} \sin \left( qL + s\Upsilon_p - \xi_m \right)
\left[ s \cos \left( qL + s\Upsilon_p + \xi_m \right) 
+ \sin \left( \kappa + \eta - s\Upsilon_m - \xi_p + 2qY_0 \right) \right]
\nonumber \\
&& \qquad - \frac{u}{2} \sin \left( qL + s\Upsilon_p + \xi_m \right)
\left[ s \cos \left( qL + s\Upsilon_p - \xi_m \right) 
- \sin \left( \kappa + \eta + s\Upsilon_m - \xi_p - 2qY_0 \right) \right] .
\label{A00NP}
\end{eqnarray}
In the degenerate case, $\xi_m = 0$, expansion of this equation for
weak potential strength up to linear in $u$ recovers the results
of the degenerate perturbation theory Eq.~(\ref{A00shifts}).
Moreover, for arbitrary potential strength and $\xi_m = 0$,
$\sin \left( qL + s\Upsilon_p \right)$
is always a common factor of Eq.~(\ref{A00NP}),
meaning that half of the levels suffer no energy shift in the presence
of an impurity for degeneracy preserving boundary conditions.
Here we are interested in the non-degenerate case where the degeneracy
has already been lifted by the boundary conditions at $y = \pm L/2$.
In the limit $u = 0$, the first term in Eq.~(\ref{A00NP}) reproduces the
results for a clean nanotube, Eqs.~(\ref{ODq1}) and (\ref{ODq2}), and we now
label these wavevectors as $q_1^{(0)}$ and $q_2^{(0)}$, respectively.
The angle $\xi_m$ breaks degeneracy,
$q_2^{(0)} - q_1^{(0)} = 2 \xi_m /L$ for $p_2 = p_1$.
Now we will present a perturbative result for weak potential strength
obtained by expanding Eq.~(\ref{A00NP}) up to linear in $u$ with $\xi_m \neq 0$:
\begin{eqnarray}
q_1 &\approx& q_1^{(0)} + \frac{su}{2L} \left[ 1 +  s (-1)^{p_{1}}
\sin \left( \kappa + \eta - s\Upsilon_m - \xi_p + 2q_1^{(0)} Y_0 \right)
\right] , 
\label{NPq1}
\end{eqnarray}
\begin{eqnarray}
q_2 &\approx& q_2^{(0)} + \frac{su}{2L} \left[ 1 - s (-1)^{p_{2}}
\sin \left( \kappa + \eta + s\Upsilon_m - \xi_p - 2q_2^{(0)} Y_0 \right)
\right] . \label{NPq2}
\end{eqnarray}
For simplicity we set $p_2 = p_1 \equiv p$ in order to show that the
impurity potential may enhance or reduce the
spacing between adjacent levels:
\begin{eqnarray}
q_2 - q_1 &=& \frac{2 \xi_m}{L} - \frac{u}{L} (-1)^{p}
\cos \left( s\Upsilon_m - 2 \left[\pi p - s \Upsilon_p \right]
\frac{Y_0}{L} \right) 
\sin \left( \kappa + \eta - \xi_p - 2 \xi_m \frac{Y_0}{L} \right) .
\end{eqnarray}
Fig.~\ref{fig:nd1} shows the evolution of energy levels near $E=0$ as a
function of the strength of the potential found by solving Eq.~(\ref{A00NP})
numerically.
In this example, there is degeneracy breaking in the clean tube due
to the boundary conditions, $\xi_m = \pi /8$.
In a similar way to the degenerate case, one of the levels in each nearly
degenerate pair does not move very much as a function of impurity strength,
while its partner suffers a shift that oscillates from pair to pair
as a function of energy (because of the non-zero position of the impurity
$Y_0$ with respect to the centre of the tube).
Some levels are brought closer together by the presence of the impurity
potential, some appear not to move, whilst others are split further apart.

\end{appendix}



\begin{thebibliography}{28}
\bibitem{saito98}  R. Saito, G. Dresselhaus, and M. S. Dresselhaus,
{\it ``Physical properties of carbon nanotubes''} (Imperial College Press,
London, 1998).

\bibitem{dek99}  C. Dekker, Physics Today {\bf 52} (5), 22 (1999).

\bibitem{bock97}  M. Bockrath, D. H. Cobden, P. L. McEuen, N. G. Chopra, A.
Zettl, A. Thess, and R. E. Smalley, Science {\bf 275}, 1922 (1997).

\bibitem{tans97}  S. J. Tans, M. H. Devoret, H. Dai, A. Thess, R. E.
Smalley, L. J. Geerlings, and C. Dekker, Nature {\bf 386}, 474 (1997).

\bibitem{ven99}  L. C. Venema, J. W. G. Wildoer, J. W. Janssen, S. J. Tans,
H. L. J. Temminck Tuinstra, L. P. Kouwenhoven, and C. Dekker, Science
{\bf 283}, 52 (1999).

\bibitem{lem01}  S. G. Lemay, J. W. Janssen, M. van den Hout, M. Mooij, M.
J. Bronikowski, P. A. Willis, R. E. Smalley, L. P. Kouwenhoven, and C.
Dekker, Nature {\bf 412}, 617 (2001).

\bibitem{lia02}
W. Liang, M. Bockrath, and H. Park, Phys. Rev. Lett. {\bf 88}, 126801 (2002).

\bibitem{bui02}
M. R. Buitelaar, A. Bachtold, T. Nussbaumer, M. Iqbal, and C. Sch\"{o}nenberger,
Phys. Rev. Lett. {\bf 88}, 156801 (2002).

\bibitem{cob02}
D. H. Cobden and J. Nyg{\aa}rd, Phys. Rev. Lett. {\bf 89}, 046803 (2002).

\bibitem{rubio99}  A. Rubio, D. Sanchez-Portal, E. Artacho, P. Ordejon, and
J. M. Soler, Phys. Rev. Lett. {\bf 82}, 3520 (1999).

\bibitem{roche99}  A. Rochefort, D. R. Salahub, and P. Avouris, J. Phys.
Chem. B {\bf 103}, 641 (1999).

\bibitem{wu00}  J. Wu, W. Duan, B.-L. Gu, J.-Z. Yu, and Y. Kawazoe, Appl.
Phys. Lett. {\bf 77}, 2554 (2000).

\bibitem{y+a01}  T.~Yaguchi and T.~Ando, J. Phys. Soc. Japan {\bf 70}, 1327
(2001); T.~Yaguchi and T.~Ando, J. Phys. Soc. Japan {\bf 70}, 3641
(2001).

\bibitem{jiang02}  J. Jiang, J. Dong, and D. Y. Xing, Phys. Rev. B {\bf 65},
245418 (2002).

\bibitem{m+f04} E. McCann and V.~I.~Fal'ko, J. Phys. Condens. Matter {\bf 16}, 2371 (2004).

\bibitem{ando98a}
T. Ando and T. Nakanishi, J. Phys. Soc. Japan {\bf 67}, 1704 (1998).

\bibitem{ando98b}
T. Ando, T. Nakanishi, and R.~Saito, J. Phys. Soc. Japan {\bf 67}, 2857 (1998).

\bibitem{ando99}
M.~Igami, T.~Nakanishi, and T.~Ando, J. Phys. Soc. Japan {\bf 68}, 716 (1999).

\bibitem{choi00}
H.~J.~Choi, J.~Ihm, S.~G.~Louie, and M.~L.~Cohen,
Phys. Rev. Lett. {\bf 84}, 2917 (2000).

\bibitem{song02}
H.-F.~Song, J.-L.~Zhu, and J.-J.~Xiong,
Phys. Rev. B {\bf 66}, 245421 (2002).

\bibitem{ke03}
S.-H. Ke, H. U. Baranger, and W. Yang,
Phys. Rev. Lett. {\bf 91}, 116803 (2003).

\bibitem{d+m84}  D. P. DiVincenzo and E. J. Mele, Phys. Rev. B {\bf 29},
1685 (1984).

\bibitem{a+a93}  H. Ajiki and T. Ando, J. Phys. Soc. Japan {\bf 62}, 1255
(1993).

\bibitem{k+m97}  C. L. Kane and E. J. Mele, Phys. Rev. Lett. {\bf 78}, 1932
(1997).

\bibitem{mceuen99}  P. L. McEuen, M. Bockrath, D. H. Cobden, Y.-G. Yoon, and
S. G. Louie, Phys. Rev. Lett. {\bf 83}, 5098 (1999).

\bibitem{del98}
P.~Delaney, H.~J.~Choi, J.~Ihm, S.~G.~Louie, and M.~L.~Cohen,
Nature {\bf 391}, 466 (1998).

\bibitem{mat98}
H. Matsumura and T. Ando, J. Phys. Soc. Japan {\bf 67}, 3542 (1998).

\bibitem{fn1}
Using ${\cal U}^{-1} H_{1d} {\cal U} = (1/2) v \Pi_z \otimes \left( \sigma_y + \sigma_z \right)
\left( -i\sigma_{y}\partial_y \right)\left( \sigma_y + \sigma_z \right)$
and $\sigma_y^2 = I_{\sigma}$ and $\sigma_z \sigma_y \sigma_z = -\sigma_y$ gives
the expression for ${\widetilde{H}_{1d}}$ shown in Eq.~(\ref{h1diag}).

\end{thebibliography}
\end{document}